# On momentum and energy of a non-radiating electromagnetic field

Alexander L. Kholmetskii
Belarusian State University
4, F. Skorina Avenue, Minsk 220080, Belarus
kholm@bsu.by

**Abstract**

This paper inspects more closely the problem of the momentum and energy of a bound (non-radiating) electromagnetic (EM) field. It has been shown that for an isolated system of non-relativistic mechanically free charged particles a transformation of mechanical to EM momentum and vice versa occurs in accordance with the requirement $\vec{P}_G$=const, where $\vec{P}_G = \vec{P}_M + \sum_{i=1}^{N} q_i \vec{A}_i$ is the canonical momentum ($N>1$ is the number of particles, $q$ is the charge, $\vec{A}$ is the vector potential, $\vec{P}_M$ is the mechanical momentum of the system). Then $\frac{d\vec{P}_M}{dt} = -\frac{d}{dt}\sum q_i \vec{A}_i$ represents the self-force, acting on this system due to violation of Newton's third law in EM interaction. If such a system contains bound charges, fixed on insulators then, according to the assumption of a number of authors, a so-called "hidden" momentum can contribute into the total momentum of the system. The problem of "hidden momentum" (*pro* and *contra*) is also examined in the paper, as well as the law of conservation of total energy for different static configurations of the system "magnetic dipole plus charged particle". Analyzing two expressions for electromagnetic momentum of a bound EM field, $\sum q_i \vec{A}_i$ and the Poynting expression $\varepsilon_0 \int_V (\vec{E} \times \vec{B}) dV$, we emphasize that they coincide with each other for quasi-static configurations, but give a discrepancy for rapid dynamical processes. We conclude that neither the first $\sum q_i \vec{A}_i$, nor the second $\varepsilon_0 \int_V (\vec{E} \times \vec{B}) dV$ expressions provide a continuous implementation of the momentum conservation law. Finally, we consider the energy flux in a bound EM field, using the Umov's vector. It has been shown that Umov vector can be directly derived from Maxwell's equations. A new form of the momentum-energy tensor, which explicitly unites the mechanical and EM masses, has been proposed.

## 1. INTRODUCTION

It is well known that local validity of the energy conservation law requires the equality of the partial time derivative of electromagnetic (EM) energy in some spatial volume $V$, $\frac{\partial}{\partial t}\int_V u\, dV$, to the energy flux across the boundary of that volume and a transmission of energy to matter. Here

$$u = \varepsilon_0 \frac{\vec{E}^2}{2} + \varepsilon_0 c^2 \frac{\vec{B}^2}{2} \tag{1}$$

is the energy density of the EM field. One sees from Eq. (1) that

$$\frac{\partial u}{\partial t} = \varepsilon_0 \vec{E} \cdot \frac{\partial \vec{E}}{\partial t} + \varepsilon_0 c^2 \vec{B} \cdot \frac{\partial \vec{B}}{\partial t}. \tag{2}$$

Considering Eq. (2), Poynting proposed to use the Maxwell equations to evaluate the field partial time derivatives [1]:

$$\frac{\partial \vec{B}}{\partial t} = -\nabla \times \vec{E}, \tag{3}$$

$$\frac{\partial \vec{E}}{\partial t} = c^2 (\nabla \times \vec{B}) - \frac{\vec{j}}{\varepsilon_0}, \tag{4}$$

Then the substitution of Eqs. (3), (4) into Eq. (2) leads to the familiar equation

$$\frac{\partial u}{\partial t} + \nabla \cdot \vec{S} + \vec{E} \cdot \vec{j} = 0 \tag{5}$$

where $\vec{j}$ is the current density, and

$$\vec{S} = \varepsilon_0 c^2 (\vec{E} \times \vec{B}) \tag{6}$$

is the Poynting vector, which defines the energy flux density of the EM field.

Applying Eqs. (5), (6) to an EM radiation, one can see that the direction of $\vec{S}$ coincides with that of EM wave propagation, and the term $\vec{j} \cdot \vec{E}$ corresponds to an absorption of EM radiation by charged particles. The same Eqs. (5) and (6) are also customarily applied to a non-radiating EM field, and according to a general theorem of classical mechanics, a momentum density $\vec{p}$ for both EM radiation and non-radiating EM field is defined as

$$\vec{p}_{EM} = \vec{S}/c^2 = \varepsilon_0 (\vec{E} \times \vec{B}). \tag{7}$$

Then the total momentum of a non-radiating EM field is computed by integration of (7) over all free space $V$:

$$\vec{P}_{EM} = \varepsilon_0 \int_V (\vec{E} \times \vec{B}) dV. \tag{8}$$

We should mention that Bessonov in a number of his papers (see, e.g. [2]) showed that the energy balance equation (5) meets a number of physical difficulties, when the point-like charged particles are involved. The problem becomes worse when the self-forces of electromagnetic fields of particles are also taken into account. However, an analysis of these problems and their resolution in [2] fall outside the scope of the present paper. In the next section we consider an isolated system of non-relativistic mechanically free charged particles and prove that a transformation of mechanical to EM momentum and vice versa occurs in accordance with the requirement $\vec{P}_G$=const, where $\vec{P}_G = \vec{P}_M + \sum_i^N q_i \vec{A}_i$ being the canonical momentum of the system ($N>1$ is the number of particles). Then $d\vec{P}_M/dt = -d/dt \left( \sum_i q_i \vec{A}_i \right)$ represents the self-force, acting on this system due to violation of Newton's third law in EM interaction. If the system contains any conductors and insulators with bound charges, a number of authors assumed that a so called "hidden momentum" $\vec{Q}_h$ should be introduced into the law of conservation of total momentum, so that $\sum_i \frac{d\vec{P}_{Mi}}{dt} = -\sum_i q_i \frac{d\vec{A}_i}{dt} - \frac{d\vec{Q}_h}{dt}$, where $\vec{Q}_h = \sum_{j=1}^{N_b} \vec{m}_{bj} \times \vec{E}_j / c^2$ is the hidden momentum, $N_b$ is the number of magnetic momenta with bound charges in the isolating system, and $\vec{E}_j$ is the electric field on the momentum $\vec{m}_{bj}$. The problem of "hidden momentum" (*pro* and *contra*) as well as the law of conservation of total energy for a static system «magnetic dipole plus charged particle» is examined in section 2. In section 3 we analyze two different expressions for electromagnetic momentum of a bound EM field, $\sum q_i \vec{A}_i$ and $\varepsilon_0 \int_V (\vec{E} \times \vec{B}) dV$, which coincide with each other for quasi-static configurations, but give a discrepancy for rapid dynamical processes. We conclude that neither the first $\sum q_i \vec{A}_i$, nor the second $\varepsilon_0 \int_V (\vec{E} \times \vec{B}) dV$ expressions provide a continuous implementation of the momentum conservation law. In section 4 we consider the energy flux in a bound EM field, using the Umov's vector. Finally, section 5 represents the conclusions.

## 2. ABOUT A MUTUAL TRANSFORMATION OF THE ELECTROMAGNETIC AND MECHANICAL MOMENTA

First consider the interaction of two non-radiating free charged particles $q_1$ and $q_2$, moving at the velocities $\vec{v}_1$ and $\vec{v}_2$ at $t=0$. One wants to determine the change with time of the total mechanical momentum of this isolating system.

It is known that the Lagrangian for a particle $q_1$ with the proper mass $m_1$ in the EM field of particle $q_2$ is

$$L_1 = -m_1 c^2 \sqrt{1 - v_1^2/c^2} - q_1 \varphi_{12} + q_1 (\vec{v}_1 \cdot \vec{A}_{12}), \qquad (9)$$

where $\varphi_{12}, \vec{A}_{12}$ are the scalar and vector potentials of the particle $q_2$ at the location of particle $q_1$. Then the motional equation of the particle $q_1$ is

$$\frac{d}{dt} \frac{\partial L_1}{\partial \vec{v}_1} = \frac{\partial L_1}{\partial \vec{r}_1}, \text{ or}$$

$$\frac{d\vec{P}_{M1}}{dt} + q_1 \frac{d\vec{A}_{12}}{dt} = -q_1 \frac{\partial \varphi_{12}}{\partial \vec{r}_1} + q_1 \frac{\partial}{\partial \vec{r}_1} (\vec{v}_1 \cdot \vec{A}_{12}), \qquad (10)$$

where $\vec{r}_1$ is the position vector of particle $q_1$, and $\vec{P}_{M1} = m_1 \vec{v}_1 / \sqrt{1 - v_1^2/c^2}$ is its mechanical momentum. In a similar way we write the Lagrangian for the particle $q_2$ with the mass $m_2$ in the field of the first particle:

$$L_2 = -m_2 c^2 \sqrt{1 - v_2^2/c^2} - q_2 \varphi_{21} + q_2 (\vec{v}_2 \cdot \vec{A}_{21}), \qquad (11)$$

where $\varphi_{21}, \vec{A}_{21}$ are the scalar and vector potentials of particle $q_1$ at the location of particle $q_2$. The motional equation is

$$\frac{d\vec{P}_{M2}}{dt} + q_2 \frac{d\vec{A}_{21}}{dt} = -q_2 \frac{\partial \varphi_{21}}{\partial \vec{r}_2} + q_2 \frac{\partial}{\partial \vec{r}_2} (\vec{v}_2 \cdot \vec{A}_{21}). \qquad (12)$$

Summing up Eqs. (10) and (12), we obtain

$$\frac{d(\vec{P}_{M1} + \vec{P}_{M2})}{dt} + \frac{d(q_1 \vec{A}_{12} + q_2 \vec{A}_{21})}{dt} = -q_1 \frac{\partial \varphi_{12}}{\partial \vec{r}_1} - q_2 \frac{\partial \varphi_{21}}{\partial \vec{r}_2} + q_1 \frac{\partial}{\partial \vec{r}_1} (\vec{v}_1 \cdot \vec{A}_{12}) + q_2 \frac{\partial}{\partial \vec{r}_2} (\vec{v}_2 \cdot \vec{A}_{21}). \qquad (13)$$

Assuming that the velocities of both particles are non-relativistic, the scalar and vector potentials produced by the particle $q_2$ at the location of particle $q_1$, and vice versa can be written to the accuracy of the order $c^{-2}$, [3]:

$$\varphi_{12} = \frac{q_2}{4\pi\varepsilon_0 r_{12}}, \quad \vec{A}_{12} = \frac{q_2\left[\vec{v}_2 + (\vec{v}_2 \cdot \hat{\vec{n}}_2)\hat{\vec{n}}_2\right]}{8\pi\varepsilon_0 c^2 r_{12}}, \quad \varphi_{21} = \frac{q_1}{4\pi\varepsilon_0 r_{21}}, \quad \vec{A}_{21} = \frac{q_1\left[\vec{v}_1 + (\vec{v}_1 \cdot \hat{\vec{n}}_1)\hat{\vec{n}}_1\right]}{8\pi\varepsilon_0 c^2 r_{21}}, \quad (14)$$

where $\vec{r}_{12} = \vec{r}_2 - \vec{r}_1$, $\vec{r}_{21} = \vec{r}_1 - \vec{r}_2$, $\hat{\vec{n}}_2$ is the unit vector at the direction from $q_2$ to $q_1$, $\hat{\vec{n}}_1$ is the unit vector from $q_1$ to $q_2$ ($\hat{\vec{n}}_2 = -\hat{\vec{n}}_1$), and $\vec{r}_1, \vec{r}_2$ are instantaneous radius-vectors of the charges $q_1$ and $q_2$. Substituting the scalar and vector potentials from Eqs. (14) into Eq. (13), one gets:

$$\frac{d(\vec{P}_{M1} + \vec{P}_{M2})}{dt} + \frac{d(q_1\vec{A}_{12} + q_2\vec{A}_{21})}{dt} = -\frac{q_1 q_2}{4\pi\varepsilon_0}\frac{\partial}{\partial \vec{r}_1}\left(\frac{1}{r_{12}}\right) - \frac{q_2 q_1}{4\pi\varepsilon_0}\frac{\partial}{\partial \vec{r}_2}\left(\frac{1}{r_{21}}\right) +$$

$$+ \frac{q_1 q_2 (\vec{v}_1 \cdot \vec{v}_2)}{8\pi\varepsilon_0 c^2}\frac{\partial}{\partial \vec{r}_1}\left(\frac{1}{r_{12}}\right) + \frac{q_2 q_1 (\vec{v}_2 \cdot \vec{v}_1)}{8\pi\varepsilon_0 c^2}\frac{\partial}{\partial \vec{r}_2}\left(\frac{1}{r_{21}}\right) +$$

$$+ \frac{q_1 q_2}{8\pi\varepsilon_0 c^2}\frac{\partial}{\partial \vec{r}_1}\left(\frac{(\vec{v}_2 \cdot \hat{\vec{n}}_2)(\vec{v}_1 \cdot \hat{\vec{n}}_2)}{r_{12}}\right) + \frac{q_2 q_1}{8\pi\varepsilon_0 c^2}\frac{\partial}{\partial \vec{r}_2}\left(\frac{(\vec{v}_1 \cdot \hat{\vec{n}}_1)(\vec{v}_2 \cdot \hat{\vec{n}}_1)}{r_{21}}\right). \quad (15)$$

Taking into account the equalities:

$$r_{12} = r_{21}, \quad \vec{r}_{12} = -\vec{r}_{21}, \quad \hat{\vec{n}}_2 = -\hat{\vec{n}}_1, \quad (16)$$

we derive

$$\frac{\partial}{\partial \vec{r}_1}\left(\frac{1}{r_{12}}\right) = -\frac{\partial}{\partial \vec{r}_2}\left(\frac{1}{r_{21}}\right), \quad (\vec{v}_2 \cdot \hat{\vec{n}}_2)(\vec{v}_1 \cdot \hat{\vec{n}}_2) = (\vec{v}_1 \cdot \hat{\vec{n}}_1)(\vec{v}_2 \cdot \hat{\vec{n}}_1),$$

$$\frac{\partial}{\partial \vec{r}_1}(\vec{v}_2 \cdot \hat{\vec{n}}_2)(\vec{v}_1 \cdot \hat{\vec{n}}_2) = -\frac{\partial}{\partial \vec{r}_2}(\vec{v}_1 \cdot \hat{\vec{n}}_1)(\vec{v}_2 \cdot \hat{\vec{n}}_1). \quad (17)$$

The obtained Eqs. (17) allow us to conclude that *rhs* of Eq.(15) is equal to zero, and

$$\frac{d(\vec{P}_{M1} + \vec{P}_{M2})}{dt} + \frac{d(q_1\vec{A}_{12} + q_2\vec{A}_{21})}{dt} = 0.$$

We can rewrite this equation as

$$\frac{d\vec{P}_G}{dt} = \frac{d\vec{P}_M}{dt} + \frac{d\vec{P}_A}{dt} = 0, \quad (18)$$

where $\vec{P}_G = \vec{P}_M + \vec{P}_A$ is the generalized (canonical) momentum, $\vec{P}_M = \vec{P}_{1M} + \vec{P}_{2M}$ is the total mechanical momentum for the isolating system of two particles, and $\vec{P}_A = q_1\vec{A}_{12} + q_2\vec{A}_{21}$. In the adopted approximation Eq. (18) is extended to the case of arbitrary number $i$ of free charged particles due to the principle of superposition:

$$\frac{d}{dt}\left(\sum_i \vec{P}_{Mi} + \sum_i q_i \vec{A}_i\right) = 0, \text{ or}$$

$$\frac{d}{dt}\sum_i \vec{P}_{Mi} = -\frac{d}{dt}\sum_i q_i \vec{A}_i. \quad (19)$$

Eq. (19) shows that the total time derivative of resultant mechanical momentum (total mechanical force, acting on the closed non-radiating system of free charged particles due to violation of Newton's third law for EM interaction) is equal with the opposite sign to the total time derivative of "momentum" $\vec{P}_A = \sum_i q_i \vec{A}_i$. Hence, under change of EM fields in the points of location of moving non-radiating particles, Eq. (19) tells us that the momentum $\vec{P}_A$ is transformed to the mechanical momentum of the non-radiating system. This assertion does not contradict the Poynting definition of EM momentum (8), because many authors proved that for finite quasi-static systems[1] of non-radiating charged particles the momentum $\sum_i q_i \vec{A}_i$ coincides with the momentum (8) (see, *e.g.*, [4, 5]). At the same time, we underline that such equality cannot be correct in a general case. Indeed, the momentum $\vec{P} = \varepsilon_0 \int_V (\vec{E} \times \vec{B}) dV$ is defined by a continuous distribution of the electric and magnetic fields over the whole free space, while the momentum $\vec{P}_A = \sum_i q_i \vec{A}_i$ is determined by the vector potential in a number of discrete spatial points $\vec{x}_i$, where the charges $q_i$ are located. Hence, for rapid dynamical processes, which essentially depend on time evolution of the EM fields and potentials ($\tau \geq L/c$), two different expressions for the momentum $\vec{P}_A = \sum_i q_i \vec{A}_i$ and $\vec{P} = \varepsilon_0 \int_V (\vec{E} \times \vec{B}) dV$ should be inevitably non-equivalent to each other. Below in section 3 we will consider a number of physical problems, which clearly indicate the difference between both momenta for dynamical EM systems.

Now look closer on physics of Eq. (19). First of all, we mention that the Lagrangian (9), used in our theorem, does not include the radiation reaction. The latter effect is taken negligible by supposition (the accelerations of particles are small)[2].

We also have to stress that Eq. (19) is valid for inertial reference frames only, although a motion of particles under observation can be arbitrary (with small accelerations, allowing for the neglect of EM radiation). Consideration of interaction between two particles from a non-inertial reference frame, attached to one of them, does not give Eq. (19) and leads to a seeming paradox with the momentum conservation law [7].

Further, we notice that the momentum $\vec{P}_A$ of a considered system is not associated with an energy flux across the boundary of that system. Following to [8], we propose to name $\vec{P}_A$ as "potential" momentum.

Eq. (19) loses its physical meaning in the case of EM radiation, when the sources of the EM field, in general, may be absent in an arbitrary space volume. Hence, for that (source-free) kind of EM fields, the momentum density is solely defined by the conventional and more general expression through the Poynting vector $\varepsilon_0 (\vec{E} \times \vec{B})$.

Eq. (19) also loses its meaning in case of a single isolating particle (*N*=1), when the term $q\vec{A}$ would mean a self-action of the particle. Hence, for such a particle the momentum of EM field is exclusively determined by Eq. (8).

Nevertheless, in many problems of classical electrodynamics, dealing with quasi-static systems, the application of "potential" momentum instead of Eq. (8) greatly simplifies calculations. In this connection it is necessary to explain that for the system of *N* free charged particles, the momentum of $i^{th}$ particle $\vec{P}_{iA} = q_i \vec{A}_i$ cannot be attributed to its proper total momentum; rather it represents a contribution of the particle *i* to the total potential momentum of the whole system; it is only $\vec{P}_A = \sum_i q_i \vec{A}_i$, which has a physical meaning. Even in the case, where occasionally the total potential momentum of a system under consideration coincides with the value $\vec{P}_{iA} = q_i \vec{A}_i$ for a single charged particle *i*, the total time derivative $-d\vec{P}_{iA}/dt$ is not equal to the force, acting on the particle *i*, but it defines the force, acting on the whole system (the charged particle *i* + the sources of the field $\vec{A}_i$). Consider, for example, the motion of a charged particle inside a mechanically free elongated solenoid[3]. Let at the initial time moment the velocity of the particle $\vec{v}$ lie in the plane *xy*, while the magnetic field of solenoid $\vec{B}$ lies in the negative *z*-direction. The Lorentz force, acting on the particle, is

$$d\vec{P}_M/dt = q(\vec{v} \times \vec{B}) = q\vec{v} \times (\nabla \times \vec{A}) = q\nabla(\vec{v} \cdot \vec{A}) - q(\vec{v} \cdot \nabla) \cdot \vec{A}.$$

---

[1] We conditionally define a quasi-static system by the requirement $\tau \gg L/c$, where $\tau$ is a typical time of dynamical processes in the system, and *L* is its typical size.
[2] For the system of radiating particles, the time derivative of momentum of free electromagnetic field should be added to *rhs* of Eq. (19) [6].

For stationary current in the solenoid, $\partial \vec{A}/\partial t = 0$, and $d\vec{A}/dt = (\vec{v} \cdot \nabla) \cdot \vec{A}$. Then, taking $\vec{P}_A = q\vec{A}$, we get

$$d\vec{P}_M/dt = -d\vec{P}_A/dt + q\nabla(\vec{v} \cdot \vec{A}). \tag{20}$$

We see that the mechanical force (the total time derivative of the momentum of particle $\vec{P}_M$) is not equal to $-d\vec{P}_A/dt$, but includes the term $q\nabla(\vec{v} \cdot \vec{A})$. However, it does not contradict Eq. (19) yet, because we did not include the mechanical momentum of solenoid $\vec{P}_{MS}$ and did not consider the force, acting on the solenoid due to the particle. One can show that this force is equal to $-q\nabla(\vec{v} \cdot \vec{A})$ (see, Appendix A), and

$$d\vec{P}_{MS}/dt = -q\nabla(\vec{v} \cdot \vec{A}). \tag{21}$$

Summing up Eqs. (20), (21), we obtain

$$d\vec{P}_M/dt + d\vec{P}_{MS}/dt = -d\vec{P}_A/dt,$$

in accordance with Eq. (19).

Let us consider another example: a charged particle *q* orbits around a tall solenoid S at the constant angular frequency $\omega$ (Appendix B, Fig. 5). In this problem the net force, acting on the particle, is equal to zero, while its "momentum" $\vec{P}_A = q\vec{A}$ changes with time. Moreover, this value defines the potential momentum of the whole system "charged particle + solenoid". Then it follows from Eq. (19) that the total time derivative $-d\vec{P}_A/dt$ should be equal to the force, acting on the solenoid due to the particle. This result is confirmed by the particular calculations, presented in Appendix B.

The revealed physical meaning of the "potential" momentum is masked in familiar textbooks, which usually begin a consideration of electrodynamics from a motion of charged particle in some abstract external EM field. By such a way it is impossible to find that the total time derivative of the momentum $\vec{P}_A = q\vec{A}$ for a given particle contains a part of force, acting on the sources of this external field. At the same time, it seems that this interpretation creates a difficulty for the energy conservation law: for example, it seems that for the problem in Appendix B the particle can rotate around the solenoid infinitely long (if we neglect its radiation), while the solenoid receives a force which can make work. This and other para-doxes (see, e.g. [5, 9]) prompted to a number of authors to introduce a so-called "hidden momentum" of a magnetic dipole. This problem is worth to be considered separately.

---

[3] Hereinafter we imagine a solenoid as two oppositely charged elongated cylinders with thin walls and equal radius, which rotate without friction at the opposite directions about a common axis with the angular frequency $\omega$. The charged are rigidly fixed on the insulating walls, that allows excluding the charge of polarization.

*2.1. "Hidden momentum": pro*

We emphasize that Eq. (19) has been derived for the isolating system of mechanically free charged particles. If the system contains any conductors, that, in general, they acquire the polarized charges with the surface charge density $\sigma_p$, which should be included into Eq. (19):

$$\frac{d}{dt}\sum_i \vec{P}_{Mi} = -\frac{d}{dt}\sum_i q_i \vec{A}_i - \frac{d}{dt}\int_S \sigma_p(\vec{r},t)\vec{A}(\vec{r},t)dS, \qquad (22)$$

where the integration is carried out over the surface $S$ of all conductors. In particular, when a conductor represents a point-like magnetic dipole $\vec{\mu}$, the integral in Eq. (22) is equal to $(\vec{\mu}\times\vec{E})/c^2$, where $\vec{E}$ is the electric field at the location of $\vec{\mu}$ [5]. The authors of ref. [5] named the value $(\vec{\mu}\times\vec{E})/c^2$ as "hidden momentum". In the present author's opinion, it is a matter of terminology solely, and we can always directly apply Eq. (22) for the charged on conductors to get correct physical results without any references on "hidden momentum".

An actual problem emerges when the system includes bound charges fixed on insulators. For such a case Shockley and James invented a paradox as follows [9].

Two counter-rotating oppositely charged insulating disks, whose rotation is slowed down by mutual friction, are in the electric field of a charged particle, which rests in a laboratory (Fig. 1). The particle and the disks lie in the plane *xy*. We want to compute the force, acting on the charged particle, as well as the force, acting on the whole isolating system "particle + rotating disks".

For the sake of simplicity we assume that the charge is homogeneously distributed over the perimeter of disks. The rotational axis of disks *z* passes across the point *x*, *y*=0, and at the initial instant the charge has coordinates {0, $R$, 0}. The radius of each disk is $r_0 < R$.

Let initially the rotational angular frequency of both disks was equal to $\omega$, and the magnetic moment $\vec{\mu}$ is parallel to the axis z. Then $\omega$ slowly decreasing to zero, so that the EM radiation is negligible. During the time $\tau$, when the frequency decreases, the vector potential of both disks also decreases with time, and induces an azimuthal electric field along the circumference $R$

$$\vec{E}(R) = -\partial\vec{A}/\partial t.$$

Taking into account that [3]

$$\vec{A} = \vec{\mu}\times\vec{R}/4\pi\varepsilon_0 c^2 R^3, \text{ and} \qquad (23)$$

$$\vec{\mu} = r_0^2 q\vec{\omega}$$

($q$ is the total charge of each disk), we derive the force, experienced by the particle $Q$:

$$F_x = QE_x = \frac{r_0^2 qQ}{4\pi\varepsilon_0 R^2}\frac{d\omega}{dt}.$$

From there the total mechanical momentum acquired by the particle $Q$ during decrease of the rotating frequency from $\omega$ to 0 is

$$(P_Q)_x = \int_0^\tau F_x dt = \frac{r_0^2 qQ}{4\pi\varepsilon_0 R^2}\int_0^\tau \frac{d\omega}{dt}dt = -QA(R). \qquad (24)$$

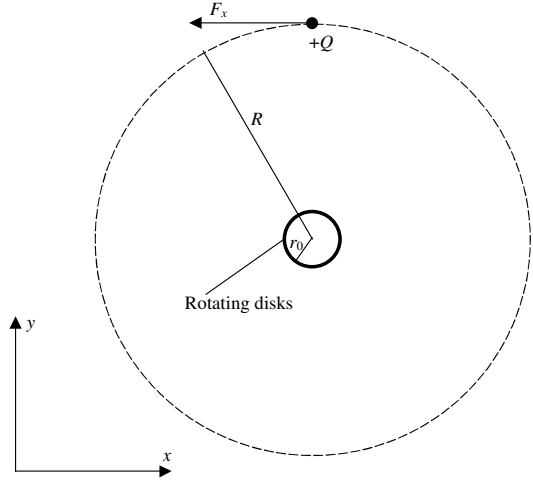

Fig. 1. The Shockley-James paradox

We see that the mechanical momentum of charged particle $Q$ after annihilation of magnetic dipole coincides with the potential momentum of the system $-QA(R)$ before annihilation. Shockley and James noticed that a motion of the particle $Q$ for resting axis of the disks means a motion of the center of mass of all system "disks + particle", which seems to contradict special relativity. In order to resolve this paradox, they introduced a "hidden" momentum of the disks $\vec{P}_h = (\vec{\mu} \times \vec{E})/c^2$, which exists due to mechanical stresses in the disks before annihilation of $\vec{\mu}$. One can check that $P_{hx} = QA(R)$, and the disks acquire the same mechanical momentum after annihilation of $\vec{\mu}$. As a result, the center of mass of the system remains at rest.

The problem of "hidden momentum" was considered in more detail by Aharonov et al [10] in connection with a classical model of neutron. As a basic point, they proved a theorem as follows:
  (a) There is zero total momentum (electromagnetic plus mechanical) in the rest frame of any finite static configuration, containing charged particles and magnetic momenta.
  (b) There is non-vanishing electromagnetic momentum of this configuration.

In order to prove part (a) of the theorem, the author used the requirement

$$\partial_\mu T^{\mu\nu} = \partial_\mu T_{EM}^{\mu\nu} + \partial_\mu T_M^{\mu\nu} = 0, \qquad (25)$$

where $\mu = 0\ldots 3$, $T_M^{\mu\nu}$ is the mechanical part of the energy momentum stress density tensor $T^{\mu\nu}$, while $T_{EM}^{\mu\nu}$ is the electromagnetic part satisfying (in MKSA units)

$$\partial_\mu T_{EM}^{\mu\nu} = -F^{\nu\lambda} j_\lambda$$

($F^{\nu\lambda}$ is the electromagnetic tensor, $j_\lambda$ is the current density). The total momentum is $P^i = \int T^{i0} dV$.

It follows from Eq. (25) that for static case $\partial_i T^{i0} = 0$ ($i=1\ldots 3$). Then one can easily prove that

$$\vec{P} = 0. \qquad (26)$$

On the other hand, proving part (b) of the theorem, the authors of [10] derived

$$\vec{P}_{EM} = \vec{E} \times \vec{\mu}/c. \tag{27}$$

One follows from Eqs. (26) and (27) that

$$\vec{P}_M = \vec{\mu} \times \vec{E}/c. \tag{28}$$

The mechanical momentum (28) represents a "hidden momentum" of the configuration, and it should be attributed to the magnetic dipole $\vec{\mu}$ exclusively. Hence, Eq. (19) should be modified as (in SI units) as

$$\sum_i \frac{d\vec{P}_{(i)}}{dt} = -\sum_i q_{(i)} \frac{\vec{A}_{(i)}}{dt} - \frac{d\vec{Q}_h}{dt}, \tag{29}$$

where $\vec{Q}_h = \sum_{j=1}^{N_b} \vec{m}_{bj} \times \vec{E}_j/c^2$ is the hidden momentum, $N_b$ is the number of magnetic momenta with bound charges in the isolating system, and $\vec{E}_j$ is the electric field on the momentum $\vec{m}_{bj}$. Then one can see that Eq. (29) fully resolves the Shockley-James paradox: the disks do indeed move in the opposite direction of the charge due to their hidden momentum, and the center of mass does remain at rest.

One should note that manifestation of hidden momentum is model dependent [10]. In the model of a magnetic dipole involving counter-rotating charged insulating disks, the external electric field causes the mechanical stresses, as mentioned in [10]. A Lorentz transformation of the stress-energy tensor converts stress to momentum density. This contribution leads to the net mechanical momentum (28) in the rest frame of the center of the disks.

*2.2. "Hidden momentum": contra*

1. A resolution of the Shockley James paradox (a rest of the center of mass) signifies that we recover the principle of equality of action and reaction for the system "charge plus magnetic dipole". One follows from there that no net external force is applied under assembling the configuration. However, this statement comes into a contradiction with the resolution of the Lewis-Tolman paradox [11].

2. A physical origin of "hidden momentum" continues to be vague. For example, for a point-like magnetic dipole, the electric field at its location can be assumed to be essentially constant. Hence, no additional mechanical stresses appear inside the dipole due to this constant field, and no converted momentum density emerges. And what is more important, a notion of "hidden momentum" contradicts the Lorentz force law, when the assembling process of the system "charge plus magnetic dipole" is considered. For example, let us analyze the problem depicted in Fig. 2. A point-like electrically neutral magnetic dipole $\vec{\mu}$ (like in Fig. 1) and a very long uniformly charged wire with the length $L$ initially are separated by a large distance $x_0$. Nevertheless, $x_0 \ll L$. Initially the dipole and wire rest in a laboratory, so that there is no interaction between them. The dipole and wire lie in the *xy*-plane, the wire is parallel to the axis *y*, and the magnetic moment $\vec{\mu}$ is parallel to the axis *z*. Then the wire acquires a very small velocity $v$ in the negative *x*-direction, and it is driven up to the distance $h$ from the dipole, while the dipole is maintained fixed in space. After this stage of assembling the velocity of wire becomes to be equal to zero, and any forces exerted on the system are vanishing. At this moment the system can be considered as static. Further, the magnetic moment slowly decreases from $\vec{\mu}$ to zero during the time $\tau$ (the stage of annihilation of magnetic dipole).

The stage of annihilation is similar to the Shockley-James paradox, and the force, acting on the wire can be found as

$$\vec{F}_w = -\int_{-\infty}^{\infty} \lambda \frac{\partial \vec{A}(h,y,t)}{\partial t} dy,$$

where $\lambda$ is the linear charge density of the wire, and $\vec{A}$ is the vector potential of magnetic dipole in the point ($x=h$; $y$; $z=0$) at the instant $t$. Then the mechanical momentum of the wire after annihilation is

$$\vec{P}_{Mw} = -\int_{t=0}^{t=\tau} \int_{-\infty}^{\infty} \lambda \frac{\partial \vec{A}(h,y,t)}{\partial t} dy dt.$$

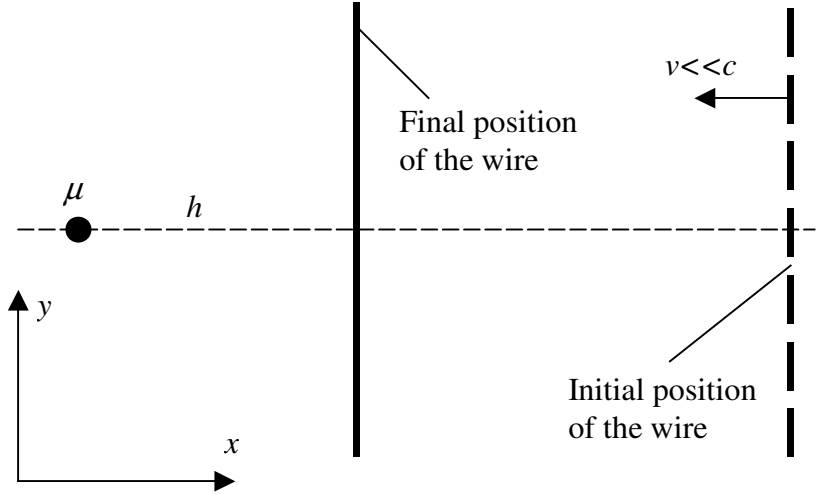

Fig. 2. Assembling of the system "magnetic dipole plus charged wire" and further annihilation of the magnetic moment $\vec{\mu}$

Insofar as the vector potential of magnetic dipole is defined by Eq. (23), we get after straight-forward calculations:

$$(P_{Mw})_x = 0, \quad (P_{Mw})_y = \frac{\lambda \mu}{2\pi \varepsilon_0 c^2 h} = \frac{\mu E}{c^2},$$

where $E = \lambda/2\pi\varepsilon_0 h$ is the value of electric field of the wire at the location of magnetic dipole. This field has only $x$-component in the plane $xy$, as we assume that the wire is very long.

According to the conception of "hidden momentum", the magnetic dipole should acquire the mechanical momentum

$$(P_{Md})_x = (\vec{\mu} \times \vec{E})_x / c^2 = 0, \quad (P_{Md})_y = (\vec{\mu} \times \vec{E})_y / c^2 = -\mu E/c^2.$$

In such a case the center of mass of the system remains at rest. One inevitably follows from there that the total mechanical momentum, transmitted to the system during its assembling, should be equal to zero. It is defined by the forces, acting on the wire and dipole during the assembling process (the motion of the wire at the constant velocity $v$ in the negative $x$-direction). While the wire is moving, it is experienced the Lorentz magnetic force

$$\vec{F}_w = \int_{-\infty}^{\infty} \lambda (\vec{v} \times \vec{B}) dy, \qquad (30)$$

where $\vec{B}$ is the magnetic field of magnetic dipole. In the plane *xy* this field has only the *z*-component

$$B_z = -\mu / 4\pi\varepsilon_0 c^2 r^3 . \tag{31}$$

Hence, the force (30) has a single component along the axis *y*:

$$(F_w)_y = -\frac{\lambda\mu v}{4\pi\varepsilon_0 c^2} \int_{-\infty}^{\infty} \frac{1}{(x^2+y^2)^{3/2}} dy = -\frac{\lambda\mu v}{2\pi\varepsilon_0 c^2 x^2} . \tag{32}$$

In order to prevent a motion of the wire along the axis *y* during the assembling of system, we have to apply to the wire a compensating counter-force

$$(F'_w)_y = \frac{\lambda\mu v}{2\pi\varepsilon_0 c^2 x^2} .$$

Then the mechanical momentum transmitted to the wire due to the compensating force is

$$(P_w)_y = \int_t (F_w')_y dt = \int_t \frac{\lambda\mu v dt}{2\pi\varepsilon_0 c^2 x^2} = \int_t \frac{\lambda\mu dx}{2\pi\varepsilon_0 c^2 x^2} = \frac{\lambda\mu}{2\pi\varepsilon_0 c^2 h} = \frac{\mu E}{c^2} . \tag{33}$$

Next compute the Lorentz force, acting on the magnetic dipole due to the moving charged wire. Since the dipole is electrically neutral, it is not experienced an electric force. In addition, the moving wire does not create a magnetic field in the plane *xy*. Indeed, in this plane the vectors $\vec{v}$ and $\vec{E}$ are collinear to each other ($\vec{v} \times \vec{E} = 0$), and no magnetic field appears. Therefore, the resultant force, acting on the magnetic dipole, is equal to zero.

The same result can be derived in another way, using the force expression [1]

$$\vec{F} = \nabla(\vec{\mu} \cdot \vec{B}) = \nabla(\mu B_z) \tag{34}$$

in the rest frame of magnetic dipole. Observe that the component of magnetic field $B_z$, created by the moving wire, is identically equal to zero, because in the proper frame of wire the component $E_y=0$ (the wire is very long), and $\vec{B}=0$. Hence, one gets from the field transformation between the wire's rest frame to the rest frame of dipole, that $B_z \equiv 0$. Thus, the force (34) is also identically equal to zero.

Finally, we can apply one more way to compute the force, acting on the magnetic dipole in the rest frame of the wire. In this frame the dipole moves at the constant velocity *v* along the *x*-axis, and an electric dipole moment $\vec{p} = \vec{v} \times \vec{\mu}/c^2$ has to appear, which the wire's electric field acts upon. This force is equal to

$$\vec{F} = (\vec{p} \cdot \nabla)\vec{E} .$$

Since only the *y*-component of $\vec{p}$ is not zero, that

$$\vec{F} = \left(p_y \frac{\partial}{\partial y}\right)\vec{E} .$$

As we mentioned above, for a long wire the electric field is not changed along its axis (the axis *y*), and this force is equal to zero.

Thus, applying the Lorentz force law, we have proved by three different ways that no force exerted on the magnetic dipole during the assembling of the system (motion of the charged wire along the axis -*x*).

From there we conclude that the mechanical momentum (33), transmitted to the charged wire during the assembling of the system, represents the total mechanical momentum, transmitted to the system at the whole. (This example explicitly indicates a violation of Newton's third law in EM interaction). Hence, a stored non-vanishing momentum of the sys-tem in its static state should inevitably exist, which coincides with the momentum (33). One can easily check that it is equal to the potential momentum of the system $\vec{P}_A = \int_{-\infty}^{+\infty} \lambda \vec{A}(\vec{r}) dy$, which, in turn, coincides with the Poynting momentum (8) for a static configuration. In fact, we observe a transformation of the mechanical momentum (33), transmitted to the system during its assembling, to the EM momentum of that system. This exact transformation of mechanical to electromagnetic momentum leaves no place for a "hidden momentum" of a magnetic dipole. Moreover, if we would demand a rest of the center of mass of the system after annihilation of magnetic dipole (causing a cancel of EM momentum), we would get a violation of the momentum conservation law: the system after annihilation has zero EM momentum and zero total mechanical momentum, while during its assembling the non-vanishing mechanical momentum (33) was transmitted to the system. Thus, there is no mystery that the center of mass of the system does move after annihilation of magnetic dipole: it occurs due to mechanical momentum, transmitted to the system during its assembling. We can name it as "retro-momentum". Such a "retro-momentum" substitutes the "hidden momentum" in the theorem by Aharonov et al. (Eq. (28)). At the same time, we have to underline that the "hidden momentum" and "retro-momentum" have quite different physical meaning. The "hidden momentum" exists instantaneously, and the center of mass of the system "magnetic dipole plus charged particle" remains at rest, while the magnetic dipole is annihilating (the Shockley-James problem); the "retro-momentum" represents the mechanical momentum transmitted to the system during its assembling in past, and stored in the EM momentum of the system. Then during annihilation of the magnetic dipole, the center of mass does move due to this stored momentum. In another words, using a notion of "retro-momentum", we inevitably get a violation of Newton's third law in EM interaction, while the introducing of "hidden momentum" maintains the equality of action and reaction in the system "charged particles plus magnetic dipoles". Hence, a rejection of "hidden momentum" seems leave the paradox in Appendix B to be non-resolved: a non-radiating charged particle, rotating around a solenoid, is not experienced any force, but induces a forced motion of the solenoid. In turn, this forced motion can make work, which becomes infinite for infinitely long rotation of the particle.

It seems that the problem of choosing between "hidden momentum" and "retro-momentum" is reduced to an unpleasant choice: either to refuse from the Lorentz force law (in favor of "hidden momentum"), or to refuse from the energy conservation law in the exercise of Appendix B (in favor of "retro-momentum"). However, next sub-section shows that the paradox of Appendix B can be resolved without "hidden momentum".

## *2.3. About momentum-energy conservation law for static configuration "charged particle and magnetic dipole (without "hidden momentum")."*

Let us analyze the energy balance for fundamental configuration "point like electrically neutral magnetic dipole plus point-like charged particle", both resting in a laboratory. The energy density of EM field is determined by Eq. (1), and the total EM energy is

$$E_{EM} = \int_V \left( \varepsilon_0 \frac{E^2}{2} + \varepsilon_0 c^2 \frac{B^2}{2} \right) dV = \frac{\varepsilon_0}{2} \int_V E^2 dV + \frac{\varepsilon_0 c^2}{2} \int_V B^2 dV. \qquad (35)$$

We see that the latter integral is divided into a sum of two parts, where the first integral (containing the electric field) is fully determined by the charge $q$ of the particle, while the second integral (containing the magnetic field) is fully determined by the magnetic moment $\vec{\mu}$ of the dipole. Therefore, for any static configuration of the system "magnetic dipole plus charged particle", the energy does not depend on the distance $\vec{r}$ between the charge and dipole, *i.e.*, it represents a constant value for fixed $q$, $\vec{\mu}$. Thus, different static configurations of the system, being characterized by different position vectors of $q$ and $\vec{\mu}$, have the same EM energy. It follows from there that any transitions between different static configurations (changes of the distances between $q$ and $\vec{\mu}$) can happen without receiving of any external energy. Indeed, under transportation of the charge $q$ from one spatial point to another at the infinitesimal velocity $\vec{v}$, the work done can be negligible, because the magnetic force, exerted by the magnetic dipole on moving charge, is always orthogonal to $\vec{v}$. During this transportation the magnetic dipole can be simply fixed in space by the external force, compensating the force due to the moving charge.

Further, one can see from Eq. (8) that the EM momentum of the system certainly depends on the distance between $q$ and $\vec{\mu}$, because it is determined by the cross product of the vectors $\vec{E}$ and $\vec{B}$. The same is true, if we define the EM momentum through the potential momentum $q\vec{A}$, $\vec{A}$ being the vector potential of the dipole at the location of charge. Thus, we conclude that different static configurations "magnetic dipole plus charged particle" represent degenerate states with respect to the energy, but non-degenerate states with respect to the momentum (for both direction and magnitude). Taking into account the law of conservation of total momentum (electromagnetic plus mechanical), we get at first glance a difficulty with implementation of the law of conservation of total energy. Indeed, under any transitions between different configurations (which occur without loss of energy), the change of EM momentum leads to the appearance of mechanical momentum of the system. As a result, a kinetic energy of the system emerges, seemingly from "nothing". How to resolve this paradox?

First of all, we have to mention that the transitions between actually static configurations imply a presence of mechanical counter-forces compensating the forces, exerted on the particle and magnetic dipole during the transition. Hence, the mechanical kinetic energy of the system remains to be equal to zero. As soon as we exclude the compensating forces, the particle and magnetic dipole both acquire non-vanishing velocities and corresponding kinetic energies during variation of $\vec{r}$. However, for these non-vanishing velocities, the magnetic dipole acquires the electric dipole moment $\vec{p} = \vec{v}_\mu \times \vec{\mu}/c^2$, while the particle produces the magnetic field. Then the magnetic dipole contributes into the electric field of Eq. (35), and charged particle contributes into the magnetic field of Eq. (35). As a result, the EM energy becomes to be dependent on $\vec{r}$. In particular, one can show that the change of kinetic energy of the system under variation of $\vec{r}$ is equal to the change of EM energy (35) with the reverse sign, and the energy conservation law is perfectly implemented.

This analysis gives a key for resolution of the paradox in Appendix B (rotation of a charged particle around a solenoid). If, by supposition, no compensating force acts on the solenoid, that it acquires a finite velocity and finite electric dipole moment. One can show that the electric field of electric dipole is opposite to the momentary velocity of particle, causing a loss of its kinetic energy. Therefore, an infinitely long rotation of the particle is impossible.

Nevertheless, a possibility to transform EM energy into kinetic energy without a loss (or minimal loss) of external energy seems very attractive, because a coefficient of efficiency of such transformer can be equal almost to unity.

## 3. MOMENTUM CONSERVATION LAW: THE REQUIREMENT OF CONTINUITY

As was mentioned above, Eq. (19)

$$\frac{d}{dt}\sum_i \vec{P}_{Mi} = -\frac{d}{dt}\sum_i q_i \vec{A}_i$$

and the Poynting expression

$$\frac{d}{dt}\sum_i \vec{P}_{Mi} = -\frac{d}{dt}\left[\varepsilon_0 \int_V (\vec{E}\times\vec{B})dV\right] \tag{36}$$

are equivalent to each other for any isolated finite quasi-static non-radiative EM system. We also emphasized that the equality of $\sum_i q_i \vec{A}_i$ and $\varepsilon_0 \int_V (\vec{E}\times\vec{B})dV$ is not valid, in general, for dynamical processes: the momentum $\varepsilon_0 \int_V (\vec{E}\times\vec{B})dV$ is defined by a continuous distribution of the electric and magnetic fields over the whole free space, while the momentum $\sum_i q_i \vec{A}_i$ is determined by the vector potential at a number of discrete spatial points $\vec{x}_i$. The non-equivalence of Eqs. (19) and (36) can be also seen from examination of their general properties.

One can see that Eq. (36) is obviously gauge-invariant. However, it is not true for Eq. (19). Indeed, the *lhs* of Eq. (19), representing the self-force, is not changed under a gauge transformation

$$A_\mu \to A'_\mu + \partial_\mu f \,,$$

($A_\mu$ is the four-potential and $f$ is an arbitrary smooth function), because it can be always expressed via the electric and magnetic fields (the Lorentz force law). At the same time, the *rhs* of Eq. (19) transforms to $\sum_i q_{(i)}\left(\dfrac{d\vec{A}'_{(i)}}{dt} + \dfrac{d}{dt}\nabla f\right)$, where, in general, $\dfrac{d}{dt}\nabla f \neq 0$ for an arbitrary smooth function $f$.

Further, Eq. (36) is the Lorentz-invariant, while Eq. (19) is not. In order to prove the latter assertion, let us try to generalize Eq. (19) into the four-dimensional form, using the four-potential and the Minkowskian force $K_\mu\{\gamma\vec{F}, \gamma(\vec{F}\cdot\vec{v})/c^2\}$. Then we get

$$\sum_i K_{(i)\mu} = -\sum_i q_{(i)} \frac{dA_{\mu(i)}}{d\tau}, \tag{37}$$

where $\tau$ is the proper time. One can see that the space-like components of Eq. (37) give Eq. (19), while its time-like component is

$$\sum_i \vec{E}_{(i)}\cdot\vec{v}_{(i)} = -\frac{d}{dt}\sum_i q_{(i)}\varphi_{(i)}$$

($\varphi$ is the electric potential), which is obviously incorrect. The correct equation is $\sum_i \vec{E}_{(i)}\cdot\vec{v}_{(i)} = -\dfrac{d}{dt}U$, where $U = \dfrac{1}{2}\sum_i q_{(i)}\varphi_{(i)}$ is the electric potential of the system of charged particles.

The non-invariance of Eq. (19) is not surprising, because it was derived on the basis of approximate expressions (14). Therefore, we may expect that not Eq. (19), but Eq. (36) should adequately describe the dynamical non-radiative systems. Unfortunately, the problem happens to be more complicated. We will show below that the requirement of a continuity of the momentum conservation law is better fitted into Eq. (19), than Eq. (36). The latter requirement implies an identical time dependence of the mechanical $\vec{P}_M$ and electromagnetic $\vec{P}_{EM}$ momenta for any isolated system, so that their sum would not depend on time:

$$\vec{P}_M(t) + \vec{P}_{EM}(t) = \text{const}. \tag{38}$$

Let us come back to the Shockley-James problem (Fig. 1), where we tacitly implied that $\tau \gg R/c$, neglecting any retardation. Now consider a limit of very large $R$, so that $\tau \ll R/c$, but the radiation processes are still negligible. In this limit we will analyze a behavior of the system at the time ranges $0<t<R/c$, $R/c<t<(R/c)+\tau$, $t>R/c+\tau$, testing both concurrent expressions for momentum of a bound EM field: Eq. (8) and "potential" momentum $Q\vec{A}$. In our analysis we omit a "hidden momentum", although its adoption only complicates the analysis, but does not essentially influence the derived conclusions. Further, we take into account a retardation effect, according to which the time-variable current $i(t)$ associated with the rotating disks produces the EM field at the instant ($t'=t-r/c$), where $r$ is the distance between the axis of point-like disks and point of observation [3].

1. $0<t<R/c$. During the time $\tau$ both disks stop rotating, and the magnetic moment vanishes. In this time range the mechanical momentum of the system is equal to zero, because a perturbation of the magnetic field from the magnetic dipole does not reach the point of location of $Q$. Correspondingly, the potential momentum of the system $\vec{P}_A = Q\vec{A}$ is also unchanged. At the same time, the Poynting momentum (8) of the disks decreases with time, because the perturbation of magnetic field is spreading in space.

2. $R/c<t<(R/c)+\tau$. In this time range the potential momentum of the system decreases from $Q\vec{A}$ to 0, and the mechanical momentum $\vec{P}_Q$ of charge $Q$ correspondingly increases from zero to $Q\vec{A}$. It fulfills the equality (38) $\vec{P}_Q(t) + \vec{P}_A(t) = const$. In this time range the momentum (8) is not changed significantly.

3. $t>R/c+\tau$. Within this time range the mechanical momentum of the system is no longer changed, while the Poynting momentum (8) continues to decrease with time, going to zero for $t\to\infty$.

The observations 1-3 clearly indicate that the law of conservation of total momentum for an isolated system is implemented continuously only in the case, where the electromagnetic momentum is defined though the potential momentum of the system. On the other hand, using the Poynting definition (8) for EM momentum, we get a violation of the momentum conservation law for all time ranges considered, excepting $t\to\infty$.

Thus, the gauge non-invariant and Lorentz non-invariant Eq. (19) has an important advantage in comparison with Eq. (36): it seems to provide a continuous validity of the momentum conservation law. However, this advantage of Eq. (19) happens to be apparent: in the next problem we will show that the conception of potential momentum, in general, does not prevent a local violation of the momentum conservation law (Fig. 3).

A magnetic dipole $\mu$ (two counter-rotating oppositely charged disks) is placed at the point $x,y,z=0$. At the initial instant the charge $+Q$ has the coordinates $\{R, 0, 0\}$, while the charge $–Q$ has the coordinates $\{R+l, 0, 0\}$, and $l\ll R$. The radius of each disk is $r_0\ll R$. Both charges and magnetic dipole are at rest in a laboratory, and there is no mutual force between them. The potential momentum of the system is equal to

$$\vec{P}_A = Q\vec{A}(R,0,0) - Q\vec{A}(R+l,0,0),$$

which, by supposition, represents the EM momentum of the system. On the axis $x$ the vector potential of the magnetic dipole is parallel to the axis $y$. Denoting this component as $A$, we get

$$P_{Ay} = Q[A(R) - A(R+l)]. \tag{39}$$

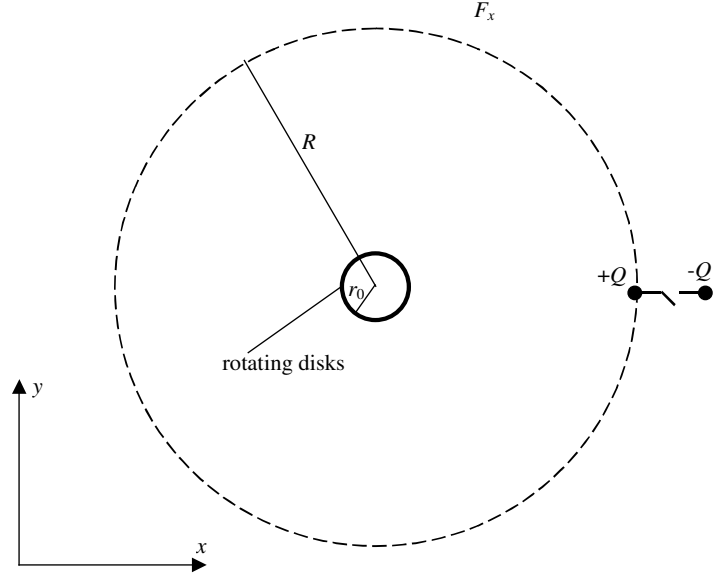

Fig. 3. The magnetic dipole from the Shockley-James problem and two opposite charged particles $+Q$ and $-Q$, which can be electrically connected with each other.

At the moment $t=0$ the charges are electrically connected by a conducting wire, and during a time interval $\tau$ a current flows in the wire. Then for $t>\tau$ both charges disappear, cancelling each other. The potential momentum (39) also vanishes. One requires determining a mechanical momentum of the system after annihilation of the charges, assuming that $\tau \ll R/c$, and the radiative EM fields are negligible. We again analyze this problem for different time intervals $0<t<\tau$, $\tau<t<(R/c)$, $R/c<t<R/c+\tau$, $t>R/c+\tau$.

1. $0<t<\tau$. At this time range a current $i(t)$ flows in the wire, which is connecting the charges $+Q$ and $-Q$. The potential momentum of the system is going to zero, and the mechanical force, acting on the wire, is equal to

$$\vec{F}_w(t) = \int_R^{R+l} [\vec{i}(t) \times \vec{B}(x)] dx, \qquad (40)$$

where $\vec{B}(x)$ is the static magnetic field of the dipole along the $x$-axis. In the plane $xy$ this field has only the $z$-component (31). Therefore, the force (40) has only a non-vanishing $y$-component. Combining Eqs. (31) and (40), we obtain

$$(F_w)_y = \frac{i(t)\mu}{4\pi\varepsilon_0 c^2} \int_R^{R+l} \frac{dx}{x^3} = \frac{i(t)\mu}{8\pi\varepsilon_0 c^2}\left(\frac{1}{R^2} - \frac{1}{(R+l)^2}\right) = \frac{i(t)[A_y(R) - A_y(R+l)]}{2},$$

where we have used Eq. (23). Hence, the total mechanical momentum, transmitted to the wire, is

$$(P_w)_y = \int_0^\tau (F_w)_y dt = \frac{[A_y(R) - A_y(R+l)]}{2} \int_0^\tau i(t) dt = \frac{Q[A_y(R) - A_y(R+l)]}{2}. \qquad (41)$$

Note that the mechanical momentum (41) is equal to the half of the potential momentum (39).

2. $\tau<t<(R/c)$. In this time interval the mechanical momentum of the magnetic dipole remains equal to zero, while the mechanical momentum of the wire is determined by Eq. (41). Thus, in this time range we observe a violation of the momentum conservation law: the total mechanical momentum of the system (41) is equal to the half of the vanished EM momentum (39).

3. $R/c < t < R/c + \tau$. Within this time interval the magnetic field $\vec{B}_w$, produced by the current $i$ in the wire, reaches the magnetic dipole. The force, experienced by the dipole, is

$$\vec{F}_d = \nabla(\vec{\mu} \cdot \vec{B}_w). \tag{42}$$

Note that the magnetic field $\vec{B}_w$ is zero at every point on the axis $x$, and in particular at the point where the dipole is, but the force is non-zero due to a non-vanishing gradient of the magnetic field. The force (42) has a single non-vanishing component on the axis $y$, and the transmitted mechanical momentum is also parallel to the axis $y$:

$$(P_d)_y = \int_{R/c}^{R/c+\tau} (F_d)_y \, dt = \int_{R/c}^{R/c+\tau} \frac{\partial}{\partial y}(\mu B_z) \, dt.$$

The simple calculations give

$$(P_d)_y = \frac{Q[A_y(R) - A_y(R+l)]}{2}. \tag{42}$$

4. $t > R/c + \tau$. Eqs. (39), (41) and (42) show that at this time range the total mechanical momentum of the configuration $(\vec{P}_w + \vec{P}_d)$ is exactly equal to the change of potential momentum $\vec{P}_A$. However, in the time ranges 1-3 the law of conservation of total momentum was temporarily violated.

Thus, the problem in Fig. 3 shows that the law of conservation of total momentum is not implemented continuously even in the case when the EM momentum is defined through a potential momentum $\vec{P}_A$. It seems that neither Eq. (19), nor Poynting's Eq. (36) satisfy the requirement (38) of continuous implementation of the total momentum conservation law. We also recall that Eq. (19) is not gauge- and Lorentz-invariant. In these conditions there is an attractive way to simultaneously resolve all problems mentioned in this section: to assume that the bound EM fields spread instantaneously in space[4]. Indeed, in this case Eq. (38) is correct for both expressions for EM momentum, and we can always use the general equation (36) as the law of conservation of total momentum. Obviously, this familiar equation is gauge- and Lorentz-invariant. At first glance, since an instantaneous action-at-a-distance can cancel special relativity, the Lorentz-invariance is no longer important. However, in alternative space-time theories, based on covariant description of ether (see, *e.g.*, [15]) the requirement of Lorentz-invariance continues to be very significant.

There is another strict physical reason to propose an instantaneous action-at-a-distance, if we consider the energy balance for the Shockley-James problem in its quasi-static and dynamical versions.

First consider a quasi-static approximation, when $\tau >> R/c$, and determine the energy of the system before and after annihilation. At the initial time moment the energy of system is composed from the mechanical rotational energy of disks $R$ and the EM energy (35). During the process of annihilation of magnetic moment, the disks receive a work W due to the frictional forces. If the disks were isolated (the charged particle is absent), then

$$W = \int_{t=0}^{t=\tau} (\vec{M}_f \cdot \vec{\omega}) \, dt, \tag{43}$$

where $\vec{M}_f$ is the torque due to the friction forces. In turn,

---

[4] We have to note that now many authors believe that "action-at-a-distance" is not forbidden by classical electrodynamics (see *e.g.* [12-14], and papers in the book mentioned in [12]).

$$W = R + \frac{\varepsilon_0 c^2}{2}\int_V B^2 dV = H, \tag{44}$$

where $\vec{B}$ is the magnetic field of the magnetic dipole before its annihilation. Here we take into account that the work of friction force $W$ is finally transformed into the heat energy $H$. Returning to the Shockley-James problem (the charged particle is present), we assume that its velocity after annihilation of the magnetic dipole is essentially non-relativistic. This allows us to neglect its magnetic field in comparison with the magnetic field of the rotating disks, as well as to take its electric energy in Eq. (35) to be constant. Hence, there is a single essential effect of a charged particle, while it is moving: the change with time of its vector potential $\partial \vec{A}_Q / \partial t$ at the location of the magnetic dipole. One sees from Fig. 1 that the induced electric field $\vec{E}_Q = -\partial \vec{A}_Q / \partial t$ exerts a torque on the disks, which is opposite to the torque due to the friction forces. It is essential that in the quasi-static limit both torques exist simultaneously (retardation is negligible). Then the expression for work received by the disks is changed in comparison with Eq. (43):

$$W' = \int_{t=0}^{t=\tau}\left((\vec{M}_f - \vec{M}_Q)\cdot\vec{\omega}\right)dt = H - \int_{t=0}^{t=\tau}(\vec{M}_Q\cdot\vec{\omega})dt = H'.$$

Thus, the moving charged particle decreases the extracted heat energy. Hence, the energy conservation law requires that

$$E_k = H - H' = \int_{t=0}^{t=\tau}(\vec{M}_Q\cdot\vec{\omega})dt. \tag{45}$$

where $E_k$ is the kinetic energy of the charged particle after annihilation of magnetic moment. This equation signifies that the particle takes its kinetic energy, reducing the heating of the disks during annihilation of magnetic moment. We omit simple but extensive calculations, which prove an exact implementation of Eq. (45).

Now consider the energy balance in the Shockley-James problem, when $\tau \ll R/c$. Then during annihilation of magnetic dipole, the charged particle $Q$ remains at rest: a perturbation of the vector potential of magnetic dipole is not yet reaching $Q$, as we assume its propagation velocity $c$. The static electric field of this charge at the location of the magnetic dipole does not create any net forces, and the initial rotational and EM energies of the disks are fully transformed into the heat $H$ (Eq. (44)). Hence, the law of conservation of energy requires that no mechanical energy should be transmitted to the particle. However, when a perturbation of magnetic field reaches the particle (at $t=R/c$), it begins to move, and at the moment ($t=R/c+\tau$) it acquires the kinetic energy $[QA(R)]^2/2M$, where $A(R)$ is the initial value of the vector potential at the location of particle. Thus, we already get a contradiction with the energy conservation law. Moreover, at the instant $t=2R/c$, a perturbation of the vector potential $\partial \vec{A}_Q / \partial t$ is reaching the charged disks, and they again begin to rotate in a direction opposite to the direction of their initial rotation at $t=0$. The friction forces slow down the disks, extracting the additional heat energy

$$\int_{t=2R/2}^{t=2R/c+\tau}(\vec{M}_Q\cdot\vec{\omega})dt = H'',$$

which should be added to the total energy of the system. As a result, we observe an obvious violation of the energy conservation law in a dynamical version of the Shockley-James problem, when the bound EM field propagates at a finite velocity. The correct energy balance equation (44) is realized, if and only if the bound EM fields spread instantaneously.

# 4. NOTES ON ENERGY FLUX IN A NON-RADIATING ELECTROMAGNETIC FIELD

We already mentioned above that the Poynting expression for energy flux density $\vec{S} = \varepsilon_0 c^2 (\vec{E} \times \vec{B})$ is traditionally applied to both free and bound EM fields, although for the latter case such energy fluxes were never detected experimentally. There is another problem with the definition $\vec{S} = \varepsilon_0 c^2 (\vec{E} \times \vec{B})$ for a bound EM field, which is revealed through its application to a single charged particle, moving at the constant velocity $\vec{v}$ in a laboratory frame.

In such a case the term $\vec{j} \cdot \vec{E}$ in Eq. (5) describes a self-action of the non-radiating inertially moving particle with its own electromagnetic field. Standard renormalization procedure implies that this term should be dropped. However, a simple cancellation of the term $\vec{j} \cdot \vec{E}$ leads to another physical difficulty. Namely, in the rest frame of a charged particle we can write ($du/dt$)=0. In the laboratory frame this equality transforms into

$$\frac{\partial u}{\partial t} + (\vec{v} \cdot \nabla) u = 0, \text{ or } \quad \frac{\partial u}{\partial t} + \nabla(\vec{v} u) = 0. \qquad (46)$$

Now let us show that Eqs. (5) and (46) are mathematically equivalent to each other, if we take into account that for the EM field of a charged particle

$$\vec{B} = \vec{v} \times \vec{E}/c^2 . \qquad (47)$$

Indeed,

$$\nabla \vec{S} = \varepsilon_0 c^2 \nabla (\vec{E} \times \vec{B}) = \varepsilon_0 c^2 [\vec{B} \cdot (\nabla \times \vec{E}) - \vec{E} \cdot (\nabla \times \vec{B})] = -\varepsilon_0 c^2 \left[ \vec{B} \cdot \frac{\partial \vec{B}}{\partial t} + \frac{\vec{E} \cdot (\nabla \times (\vec{v} \times \vec{E}))}{c^2} \right] =$$

$$= -\varepsilon_0 c^2 \left[ \vec{B} \cdot \frac{(\vec{v} \times \partial \vec{E}/\partial t)}{c^2} + \frac{\vec{E} \cdot [\vec{v} \cdot (\nabla \cdot \vec{E})]}{c^2} - \frac{\vec{E} \cdot [(\vec{v} \cdot \nabla) \cdot \vec{E}]}{c^2} \right] = \qquad (48)$$

$$= -\varepsilon_0 \vec{B} \cdot (\vec{v} \times \partial \vec{E}/\partial t) - \vec{E} \cdot \vec{j} + \varepsilon_0 \vec{E} \cdot [(\vec{v} \cdot \nabla) \cdot \vec{E}]$$

Here we used the vector identity $\vec{a} \times (\vec{b} \times \vec{c}) = \vec{b} \cdot (\vec{a} \cdot \vec{c}) - \vec{c} \cdot (\vec{a} \cdot \vec{b})$, as well as the equality

$$\vec{v} \cdot (\nabla \vec{E}) = \vec{v} \frac{\rho}{\varepsilon_0} = \frac{\vec{j}}{\varepsilon_0},$$

where $\rho$ is the charge density. Further, using Eq. (4), we can write

$$\vec{B} \cdot (\vec{v} \times \partial \vec{E}/\partial t) = \vec{B} \cdot \left[ \vec{v} \times \left( c^2 (\nabla \times \vec{B}) - \frac{\vec{j}}{\varepsilon_0} \right) \right] = -c^2 \vec{B} \cdot [(\vec{v} \cdot \nabla) \cdot \vec{B}] \qquad (49)$$

(under transformation of Eq. (49) we again use the identity $\vec{a} \times (\vec{b} \times \vec{c}) = \vec{b} \cdot (\vec{a} \cdot \vec{c}) - \vec{c} \cdot (\vec{a} \cdot \vec{b})$, and take into account that the vectors $\vec{v}$ and $\vec{B}$ are orthogonal to each other, so that $\vec{v} \cdot \vec{B} = 0$. From Eqs. (49) and (48) we derive

$$\nabla \vec{S} = \varepsilon_0 \vec{E} \cdot [(\vec{v} \cdot \nabla) \cdot \vec{E}] + \varepsilon_0 c^2 \vec{B} \cdot [(\vec{v} \cdot \nabla) \cdot \vec{B}] - \vec{E} \cdot \vec{j} . \qquad (50)$$

Substituting $\nabla \vec{S}$ given by Eq. (50) into Eq. (5), we obtain:

$$\frac{\partial u}{\partial t} + \varepsilon_0 \vec{E} \cdot [(\vec{v} \cdot \nabla) \cdot \vec{E}] + \varepsilon_0 c^2 \vec{B} \cdot [(\vec{v} \cdot \nabla) \cdot \vec{B}] = 0 . \qquad (51)$$

In turn, one can see that

$$\varepsilon_0 \vec{E} \cdot [(\vec{v} \cdot \nabla) \cdot \vec{E}] + \varepsilon_0 c^2 \vec{B} \cdot [(\vec{v} \cdot \nabla) \cdot \vec{B}] = \nabla \left[ \vec{v} \left( \frac{\varepsilon_0 E^2}{2} + \frac{\varepsilon_0 c^2 B^2}{2} \right) \right] = \nabla (\vec{v} u),$$

and Eq. (51) transforms to Eq. (46). We can rewrite Eq. (46) as

$$\frac{\partial u}{\partial t} + \nabla \vec{S}_U = 0, \quad (52)$$

where

$$\vec{S}_U = \vec{v} u \quad (53)$$

is known as Umov's vector [17].

Thus, for a charged particle, moving at a constant velocity $\vec{v}$, we derived two mathematically equivalent forms of the energy balance equation:

$$\frac{\partial u}{\partial t} + \nabla \vec{S} + \vec{j} \cdot \vec{E} = 0, \text{ and } \quad \frac{\partial u}{\partial t} + \nabla \vec{S}_U = 0.$$

One sees from there that we cannot simply omit the term $\vec{j} \cdot \vec{E}$ in Eq. (5), because by such a way we destroy an equivalence of Eqs. (5) and (46). From this point of view Eq. (46) seems more attractive, because it does not contain this term of self-action. Physically this means that the electromagnetic field moves uniformly in space together with its charged particle source at the velocity $\vec{v}$. On the other hand, the EM field in Maxwell's electrodynamics "knows" only two velocities: 0 and $c$. Hence, a representation of the EM field, moving at $v<c$ cannot describe a real physical situation within this theory. As a result, we reveal that Eqs. (5) and (46), being equivalent mathematically, are both physically unsatisfactory for the Maxwell theory in the description of the energy flux of a single charged particle. In these conditions, the following method of overcoming this difficulty has been proposed (see [18] and references therein): both expressions for the energy flux density, $\vec{S}$ and $\vec{S}_U$ are relevant, but the first of them describes "differential energy fluxes" $\vec{S}$, propagating at the velocity $c$, which compose an "integral energy flux" $\vec{S}_U$, propagating with an effective velocity equal to the velocity of the source particle $\vec{v}$.

However, this idea seems to be wrong, when the system of $N>1$ charged particles is considered. Let us denote $\vec{v}_l$ the instantaneous velocities of particles at the considered instant ($l=1...N$). We assume that there are no external mechanical forces, and the accelerations of particles are small enough to neglect their EM radiation. Generalizing the calculations (47)-(51) to this system, we again obtain Eq. (52), where $\vec{S}_U$ is substituted by the vector

$$\vec{S}_{UG} = \varepsilon_0 \sum_l \vec{v}_l \frac{(\vec{E}_\Sigma \cdot \vec{E}_l)}{2} + \varepsilon_0 c^2 \sum_l \vec{v}_l \frac{(\vec{B}_\Sigma \cdot \vec{B}_l)}{2}. \quad (54)$$

Here $\vec{E}_\Sigma = \sum_l \vec{E}_l$, $\vec{B}_\Sigma = \sum_l \vec{B}_l$ are the resultant electric and magnetic fields created by the charged particles. We can name the vector $\vec{S}_{UG}$ in Eq. (54) as "generalized" Umov vector, and it describes the "integral" energy flux density of a system of charged particles. We underline that Eq. (54) is compatible with Maxwell's equations, because it has been obtained under generalization of Eqs. (47)-(52) to the case of $N$ charged particles. If all particles move uniformly at the same momentary velocity $\vec{v}$, the generalized Umov vector becomes equal to

$$\vec{S}_{UG} = \vec{v} \varepsilon_0 \frac{1}{2} \left( \vec{E}_\Sigma^{\,2} + c^2 \vec{B}_\Sigma^{\,2} \right). \quad (55)$$

This shows that the resultant fields again move uniformly with the system of charged source particles.

Now let us demonstrate that Eq. (55) disproves the idea about "differential" and "integral" energy fluxes. Indeed, consider the motion of a charged parallel plate capacitor in the direction normal to the plates (Fig. 4). The square plates have the size $a \times a$, where $a \gg h$, $h$ being the distance between the plates. Then in the space region far from the boundaries of the plates, the electric field $\vec{E}$ is constant and coincides with the direction of velocity of the plates $\vec{v}$. Since the magnetic field is absent between the plates ($\vec{v} \times \vec{E} = 0$), then the generali-zed Umov vector is equal to $\vec{S}_{UG} = \vec{v}\varepsilon_0 E^2/2$: the electric field rigidly moves together with the plates. However, by no way can this result be understood with the Poynting vector. Indeed, in this space region $\vec{S} = \varepsilon_0 c^2 (\vec{E} \times \vec{B}) = 0$, and there is no energy flux inside the capacitor in the Poynting's meaning. Therefore, the mentioned above conception about "differential" and "integral" energy flux cannot describe this effect. Only Eq. (55) remains relevant, and it should be interpreted in its direct meaning: the EM field rigidly moves together with the source particles. Hence, the same interpretation must be applied to the single particle (Eq. (53). Insofar as the velocity of the particle exactly coincides with the velocity of the EM field configuration in the whole space, we have to adopt an instantaneous propagation of the bound EM fields.

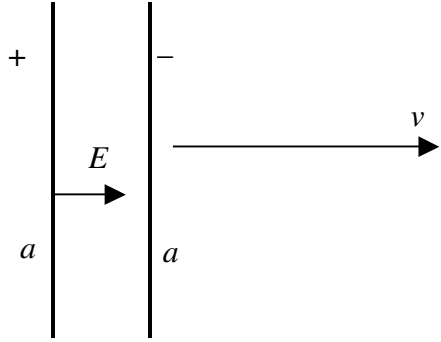

Fig. 4. A parallel plate charged capacitor moves at the constant velocity v along the normal to the plates.

We again underline that Eqs. (52)-(55), leading to the instantaneous action-at-a-distance, simultaneously represent the solutions of Maxwell's equations. In this connection it is worth to mention the plurality of solutions of Maxwell's equations [13, 19] revealed during past years, in particular, co-existence of hyperbolic and elliptic solutions. The latter correspond to an instantaneous spread in space of bound EM fields, guided by the generalized Umov vector.

These results indicate that free and bound EM fields represent different physical entities. Indeed, the non-radiating EM fields obey non-homogeneous Maxwell's equation, while EM radiation obeys homogeneous Maxwell equations. It simultaneously means that there should be a physical mechanism, allowing distinguishing the bound and free EM fields in their mixture. In the author's opinion, such a physical mechanism is based on the fact that EM radiation is absorbed, at least in principle, by a charged particle, while a non-radiating field is not (the change of kinetic energy of particle is equal to change of its electric potential energy with the reverse sign). Therefore, such a difference between free and bound EM fields should be reflected in the energy-momentum tensor $T_{EM}{}^{\mu\nu}$. Let us closer explore this problem.

It is known that the motional equation for an EM field with the Lagrangian density $-\frac{\varepsilon_0}{4}F_{\mu\nu}F^{\mu\nu}$ ($F^{\mu\nu}$ is the tensor of EM field) gives [3]

$$T_{EM}{}^{\mu\nu} = -\varepsilon_0 \partial_\mu A^\gamma F_\gamma^\nu + \frac{\varepsilon_0}{4} g^{\mu\nu} F_{\gamma\alpha} F^{\gamma\alpha}. \tag{56}$$

where **g** is the metric tensor. A physically meaningful energy-momentum tensor should be symmetrical. Using the gauge arbitrariness in its choice,

$$T_{EM}{}^{\mu\nu} \to T_{EM}{}^{\mu\nu} + \partial_\gamma \psi^{\mu\nu\gamma} \text{ (where } \psi^{\mu\nu\gamma} = \psi^{\mu\gamma\nu}\text{)}, \tag{57}$$

the tensor (56) can be transformed to the symmetric form

$$T_{EM}{}^{\mu\nu} = \varepsilon_0 \left( -F^{\mu\gamma} F_\gamma^\nu + \frac{1}{4} g^{\mu\nu} F_{\gamma\alpha} F^{\gamma\alpha} \right). \tag{58}$$

Eq. (57) represents the conventional expression for the tensor of EM field. One should recall that the transformation of Eq. (56) into Eq. (58) uses the equality [3]

$$\partial_\mu F^{\mu\nu} = 0, \tag{59}$$

which represents the Maxwell equation for a source-free EM field. This fact (the gauge function $\psi$ for transformation of Eq. (57) into (58) is determined for source-free EM fields only, while the tensor (58) is considered as general) was dropped without any comments. Below we will find a physical meaning for the gauge transformation with the condition (59).

It is known that the components $T_{EM}{}^{i0}$ of the tensor (58) determine the Poynting vector $S^i$. Now let us find the form of an energy-momentum tensor, whose components $T_{EM}{}^{i0}$ compose the generalized Umov vector. One of the methods to solve this problem is to test different gauge functions $\psi^{\mu\nu\gamma}$ in (57) to obtain the components $T_{EM}{}^{i0}$, given by Eq. (54). We can avoid such a complex way, using the requirement

$$\partial_\mu T_{EM}{}^{\mu\nu} = 0, \tag{60}$$

for free space volume. Then in my earlier paper [16] the following tensor of EM energy has been suggested, satisfying both conditions (54) and (60):

$$T^{\mu\nu} = \frac{1}{2} \sum_l \frac{dx^\mu{}_{(l)}}{dt} \frac{dx^\nu{}_{(l)}}{dt} \left[ \varepsilon_0 (\vec{E}_\Sigma, \vec{E}_l) + \varepsilon_0 c^2 (\vec{B}_\Sigma, \vec{B}_l) \right]. \tag{61}$$

However, this equation is not exactly correct. Indeed, the term $\left[ \varepsilon_0 (\vec{E}_\Sigma, \vec{E}_l) + \varepsilon_0 c^2 (\vec{B}_\Sigma, \vec{B}_l) \right] / c^2$ has the dimension of mass density, and its relativistic dependence on velocity should be included. It can be done for the following modification of the tensor, which also satisfies the requirements (54) and (60):

$$T^{\mu\nu} = \frac{1}{2c^2} \sum_l \frac{dx^\mu{}_{(l)}}{dt} \frac{dx^\nu{}_{(l)}}{d\tau} \left[ \varepsilon_0 (\vec{E}_\Sigma, \vec{E}_l) + \varepsilon_0 c^2 (\vec{B}_\Sigma, \vec{B}_l) \right] \tag{62}$$

where $\tau$ is the proper time. Simultaneously we use a conventional definition $x^0 = ct$, introducing the multiplier $1/c^2$.

Obviously, the tensors (58) and (62) differ from each other, and this puzzling was not resolved in [16]. Now we can establish a relationship between both tensors and propose a new form of the energy-momentum tensor.

Consider again the system of *N* charged particles. For each $l^{th}$ particle ($l=1…N$) we distinguish its own bound EM field and the external EM fields of other ($N-1$) particles. Then at the location of the particle $l$, $\left(\partial_\mu F^{\mu\nu}\right)_{external(l)} = 0$. Now we see that it is just the condition (59). Hence, it is naturally to take the tensor (58) for description of all external EM fields (both free and bound) at the location of particle *l*. In general, it also includes EM radiation of this particle. A remaining step is to introduce the bound EM field (62) of $l^{th}$ particle into the tensor of EM field. Then the total energy-momentum tensor can be written as the sum

$$T^{\mu\nu} = T_M^{\mu\nu} + T_{EM(bl)}^{\mu\nu} + T_{EMex(l)}^{\mu\nu}, \tag{63}$$

where $T_M^{\mu\nu}$ is the mechanical energy-momentum tensor [3]

$$T_M^{\mu\nu} = \sum_l \mu_{(l)} \frac{dx^\mu{}_{(l)}}{dt} \frac{dx^\nu{}_{(l)}}{d\tau}, \tag{64}$$

$\mu_{(l)} = \mu(\vec{r}_l, t)$ being the mass density in the point $\vec{r}_l$. The subscript (*bl*) in Eq. (63) denotes the bound EM field of $l^{th}$ particle, and the subscript "ex(*l*)" designates all EM fields in the system, excepting bound EM field of $l^{th}$ particle. Combining Eqs. (64), (63), (62) and (58), we derive the explicit form of the total energy-momentum tensor:

$$T^{\mu\nu} = \sum_l \left\{ \left( \mu_{(l)} + \frac{\varepsilon_0}{2c^2}\left[\left(\vec{E}_\Sigma, \vec{E}_l\right) + c^2\left(\vec{B}_\Sigma, \vec{B}_l\right)\right] \right) \frac{dx^\mu{}_{(l)}}{dt} \frac{dx^\nu{}_{(l)}}{d\tau} + \varepsilon_0 \left( -F^{\mu\gamma} F^\nu{}_\gamma + \frac{1}{4} g^{\mu\nu} F_{\gamma\alpha} F^{\gamma\alpha} \right)_{ex(l)} \right\}$$

(65).

Let us introduce the total mass density

$$\mu_{t(l)} = \mu_{(l)} + \frac{\varepsilon_0}{2c^2}\left[\left(\vec{E}_\Sigma, \vec{E}_l\right) + c^2\left(\vec{B}_\Sigma, \vec{B}_l\right)\right], \tag{66}$$

where the term $\frac{\varepsilon_0}{2c^2}\left[\left(\vec{E}_\Sigma, \vec{E}_l\right) + c^2\left(\vec{B}_\Sigma, \vec{B}_l\right)\right]$ should be interpreted as the mass density of a bound EM field $\mu_{EM}$. Then one can prove that the tensor (65) satisfies the law of conservation of total momentum-energy, $\partial_\mu T^{\mu\nu} = 0$, if and only if the motional equation for $l^{th}$ non-radiating charged particle contains not the mechanical, but the total mass $m_t$:

$$m_{t(l)} \left(\frac{du_\mu}{d\tau}\right)_{(l)} = q_{(l)} \left(F_{\mu\nu}\right)_{ex(l)} u^\nu, \tag{67}$$

where $u_\mu$ is the four-velocity, and $q_{(l)}$ is the charge of particle *l*. Eq. (67) shows that "mechanical" and "electromagnetic" masses are both the intrinsic parts of a total mass of charged particle. The idea to include the EM mass in the total mass of charged particles is as old as the classical model of the electron. However, to the best of the author's knowledge, this idea was forgotten every time, when the tensor of EM field and the motional equation were derived. Thus, the corrected Eq. (67) seems significant. We notice that its *rhs* contains the tensor $F_{\mu\nu}$, belonging to the external EM fields, and the self-action is excluded.

It is well recognized that within classical electrodynamics we cannot distinguish the relative contributions of mechanical and EM masses into the total mass $m_t$, which is a measurable value only. Therefore, for practical purposes we can omit the subscript "t", and simply designate the measurable mass as *m*. Thus, we see that introducing into the total energy momentum tensor of the component (62) (related with the generalized Umov vector (54)), only recalls us that the parameter "*m*" in the motional equation represents the sum of mechanical and electromagnetic masses, and, what is more important, excludes the force of self-action.

At the same time, the obtained energy-momentum tensor (65) essentially influences our understanding of the energy fluxes in EM fields.

Indeed, first consider free of source particles space. One sees that the tensor (65) is reduced to the tensor (58), which determines the Poynting vector $\vec{S}$. Then we conclude that propagation of EM radiation is given by $\vec{S}$, the well-known result of classical electromagnetism.

Now consider a bound EM field of a system of $N$ charged particles. Then both terms in Eq. (65) are relevant, and it seems at first glance that both Umov and Poynting vectors are relevant for description of the energy fluxes in such EM field. However, it is not correct: the energy fluxes in bound EM fields are determined by the Umov vector solely. We demonstrate the validity of this assertion with a well-known problem: the energy flux in an EM field of a straight wire with steady current $I$ (Appendix C).

Thus, we conclude that free and bound EM fields actually represent different physical entities: the free fields are guided by Poynting vector, while the bound fields are guided by Umov vector. There is another attractive point in the understanding, that the generalized Umov vector (54) describes propagation in space of a bound EM field and its mass density. For a single moving particle, Eq. (54) transforms to the conventional Umov vector (53), which shows that EM mass moves at the same velocity as a source particle. The momentum density of a propagating bound EM field is determined as

$$\vec{p}_{EM} = \frac{\vec{S}_u}{c^2} = \vec{v}\frac{u}{c^2} = \mu_{EM}\vec{v}. \qquad (68)$$

One sees that Eq. (68) provides the equality

$$u = \mu_{EM} c^2 \qquad (69)$$

for the energy and mass densities in accordance with the Einstein's expression. Integrating Eq. (69) over the whole space, we get

$$U_{EM} = M_{EM} c^2$$

($U_{EM}$ is the total EM energy, $M_{EM}$ is the total EM mass). Thus, the expression of energy flux density through the vector of Umov eliminates the familiar problem "4/3" [20] for a moving electron, and realizes the relativistic expression (69) for the mass and energy densities.

## 5. CONCLUSIONS

1. It has been shown that for an isolated system of non-relativistic mechanically free charged particles, its canonical momentum $\sum_i (\vec{P}_{Mi} + q_i \vec{A}_i)$ is conserved. Hence, $\frac{d}{dt}\sum_i \vec{P}_{Mi} = -\frac{d}{dt}\sum_i q_i \vec{A}_i$ describes the self-force, acting on this system due to violation of Newton's third law in EM interaction. If such a system contains bound charges, fixed on insulators, then a "hidden" momentum can (by supposition of a number of authors) contribute to the total momentum of the system. However, we have shown that the conception of "hidden momentum" leads to a contradiction with the Lorentz force law (the problem in Fig. 2). Rejecting a "hidden momentum" and introducing a "retro-momentum" instead, we reveal that a static system "magnetic dipole plus charged particle" has a property as follows: its different configurations, characterized by different position vectors of the dipole and particle, represent degenerate states with respect to EM energy, but non-degenerate states with respect to EM momentum (for both direction and magnitude). This opens a possibility to transform EM energy into kinetic energy without a loss (or minimal loss) of external energy.

2. The introduced "potential momentum" of an EM system $\vec{P}_A = \sum_i q_{(i)} \vec{A}_{(i)}$ and conventional momentum of a bound EM field $\vec{P}_{EM} = \varepsilon_0 \int_V (\vec{E} \times \vec{B}) dV$ are equal to each other only for quasi-static configurations. For rapid dynamical processes, where a time evolution of bound EM fields is essential, the two expressions for EM momentum are not equivalent to each other, and neither provides a continuous implementation of the total momentum conservation law, if the bound EM fields propagate at the velocity $c$. Moreover, we found that the energy conservation law cannot be fulfilled for the dynamical system "magnetic dipole plus charged particle", if the bound EM fields propagate at finite velocity. Hence, an introducing of instantaneous action-at-a-distance is strongly indicated.

3. It has been shown that the generalized Umov's vector, describing an energy flux density for the bound EM fields, can be obtained from Maxwell's equations. A new form of the momentum-energy stress density tensor has been proposed. A four-divergence of this tensor is vanishing, if and only if a mass of bound EM field is included in the total mass of source particles. Simultaneously the forces of self-action are excluded from the motional equation. A bound EM field (and associated mass) rigidly moves in space together with a source particle, which is possible only for elliptic solutions of non-homogeneous Maxwell's equations (instantaneous action-at-a-distance). Such a source-bound EM field might be represented quantum mechanically as a cloud of virtual photons not subject to causal limitations.

**Acknowledgments**

The author warmly thanks Thomas E. Phipps, Jr., for the discussion of some points of this paper.

**Appendix A. Calculation of the force, acting on a solenoid due to a charged particle, moving inside the solenoid**

Let at the initial instant $t=0$ the charged particle $q$ moves inside a tall solenoid at the velocity $\vec{v}$, lying in the $xy$-plane (Fig. 5). The radius of the solenoid is equal to $r$, the distance between the particle and the axis of solenoid is $R<r$ at $t=0$. The axis of solenoid is collinear to the axis $z$. One requires to determine the force, experienced by the solenoid, carrying the current $i$.

In the non-relativistic limit a moving charged particle creates the magnetic field

$$\vec{B} = \frac{q\vec{v} \times \hat{\vec{n}}}{4\pi\varepsilon_0 c^2 R'^2}, \qquad (A1)$$

where $\hat{\vec{n}}$ is the unit vector, joining the point-like charge and designated space point $R'$. This field induces a magnetic force, acting on each element $dl$ of solenoid with the current $\vec{i}$:

$$d\vec{F} = (\vec{i} \times \vec{B}) dl. \qquad (A2)$$

Without a lose of generality, we can choose the axis $x$ to be orthogonal to $\vec{v}$ at $t=0$. Then the component $B_y=0$ in Eq. (A1). Also taking into account, that $i_z=0$ in the solenoid, we obtain the components of force for a single loop of the solenoid as

$$dF_{lx} = i_y B_z dl = -i B_z r \cos\varphi\, d\varphi, \tag{A3}$$

$$dF_{ly} = -i_x B_z dl = -i B_z r \sin\varphi\, d\varphi, \tag{A4}$$

$$dF_{lz} = -i_y B_x dl = i B_x r \cos\varphi\, d\varphi. \tag{A5}$$

where $\varphi$ is the circumferential angle (Fig. 5, a). Firstly, let us calculate the component of total force along the axis $x$. Eq. (A1) gives:

$$B_z = -\frac{qv n_x}{4\pi\varepsilon_0 c^2 R'^2}. \tag{A6}$$

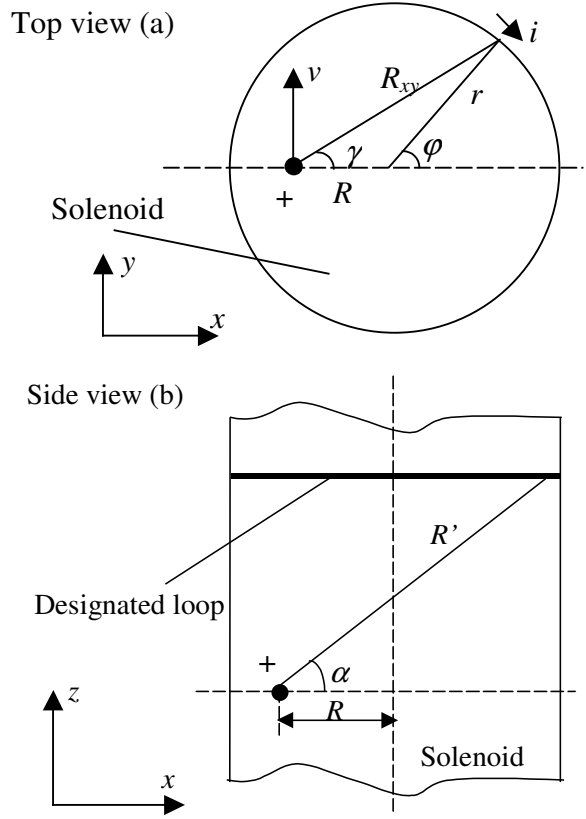

Fig. 5. The charged particle +q moves inside the solenoid. We imagine a solenoid as two oppositely charged insolating elongated cylinders with thin walls and equal radius, which rotate without friction at the opposite directions about a common axis and the angular frequency $\omega$. In order to apply the same expressions, like for conductive solenoid, we assume that each cylinder contains N equally charged layers with the charge Q and with n layers per unit length. Then the current in each layer is equal to $i=Q\omega/2\pi$.

Substituting Eq. (A6) into Eq. (A3), we obtain

$$dF_{lx} = \frac{qv n_x i r \cos\varphi\, d\varphi}{4\pi\varepsilon_0 c^2 R'^2}. \tag{A7}$$

One can see from Fig. 5, that

$$R'^2 = R_{xy}^2 + z^2,\ R_{xy}^2 = R^2 + 2Rr\cos\varphi + r^2,\ \cos\alpha = \frac{R_{xy}}{R'},\ \cos\gamma = \frac{R + r\cos\varphi}{R_{xy}},$$

$$n_x = \cos\alpha \cos\gamma = \frac{R + r\cos\varphi}{\sqrt{R^2 + 2Rr\cos\varphi + r^2 + z^2}}. \tag{A8}$$

Substituting the values of (A8) into (A7), one gets:

$$dF_x = \frac{qvir(R+r\cos\varphi)\cos\varphi d\varphi}{4\pi\varepsilon_0 c^2 (R^2 + 2Rr\cos\varphi + r^2 + z^2)^{3/2}}. \tag{A9}$$

From there the force, acting on a single loop of solenoid along the axis $x$, is

$$dF_{lx} = \frac{qvir}{4\pi\varepsilon_0 c^2} \int_0^{2\pi} \frac{(R+r\cos\varphi)\cos\varphi d\varphi}{(R^2 + 2Rr\cos\varphi + r^2 + z^2)^{3/2}} = \frac{qvi}{4\pi\varepsilon_0 c^2} \int_0^{2\pi} \frac{\left(\frac{R}{r}+\cos\varphi\right)\cos\varphi d\varphi}{\left(1+\frac{2R}{r}\cos\varphi + \frac{R^2}{r^2} + \frac{z^2}{r^2}\right)^{3/2}}.$$

The fragment of solenoid with the length $dz$ contains $ndz$ loops. Hence, the force, acting on this fragment is

$$dF_{lx} = \frac{qvindz}{4\pi\varepsilon_0 c^2} \int_0^{2\pi} \frac{\left(\frac{R}{r}+\cos\varphi\right)\cos\varphi d\varphi}{\left(1+\frac{2R}{r}\cos\varphi + \frac{R^2}{r^2} + \frac{z^2}{r^2}\right)^{3/2}}.$$

From there the total force along the axis $x$, acting on the solenoid due to the moving particle, is

$$F_x = \frac{qvin}{4\pi\varepsilon_0 c^2} \int_{-\infty}^{\infty}\int_0^{2\pi} \frac{\left(\frac{R}{r}+\cos\varphi\right)\cos\varphi d\varphi dz}{\left(1+\frac{2R}{r}\cos\varphi + \frac{R^2}{r^2} + \frac{z^2}{r^2}\right)^{3/2}}.$$

Taking into account that $in/\varepsilon_0 c^2 = B$, we obtain

$$F_x = \frac{qvB}{4\pi} \int_{-\infty}^{\infty}\int_0^{2\pi} \frac{\left(\frac{R}{r}+\cos\varphi\right)\cos\varphi d\varphi dz}{\left(1+\frac{2R}{r}\cos\varphi + \frac{R^2}{r^2} + \frac{z^2}{r^2}\right)^{3/2}}.$$

Integration over $z$ gives:

$$F_x = \frac{qvB}{4\pi} \int_0^{2\pi}\left(\frac{R}{r}+\cos\varphi\right)\cos\varphi d\varphi \int_{-\infty}^{\infty} \frac{dz}{\left(1+\frac{2R}{r}\cos\varphi + \frac{R^2}{r^2} + \frac{z^2}{r^2}\right)^{3/2}} =$$

$$= \frac{qvB}{2\pi} \int_0^{2\pi} \frac{\left(\frac{R}{r}+\cos\varphi\right)}{\left(1+\frac{2R}{r}\cos\varphi + \frac{R^2}{r^2}\right)}\cos\varphi d\varphi. \tag{A10}$$

The remaining integral over $\varphi$ is equal to

$$\int_{-\infty}^{\infty} \frac{\left(1+\frac{r}{R}\cos\varphi\right)}{1+\frac{2r}{R}\cos\varphi + \frac{r^2}{R^2}}\cos\varphi d\varphi = \pi. \tag{A11}$$

Substituting Eq. (A11) into Eq. (A10), we obtain

$$F_x = qvB/2. \tag{A12}$$

Taking into account that inside the solenoid the vector potential is equal to $A = BR/2$ and circulated in the clock-wise direction, one sees that Eq. (A12) gives the same force, as

$$F_x = -\left[q\nabla(\vec{v}\cdot\vec{A})\right]_x$$

(see, Eq. (21)). In a similar way one can show that the components $F_y$ and $F_z$, computed from Eqs. (A4) and (A5), coincide with corresponding components of the force

$$\vec{F} = -q\nabla(\vec{v}\cdot\vec{A}). \tag{A13}$$

Thus, the moving charged particle creates the force (A13), exerted on the solenoid.

## Appendix B. Calculation of the force, acting on a solenoid due to a charged particle, rotating around the solenoid

Let a charged particle $q$ orbits in the $xy$-plane around a tall solenoid S at the constant angular frequency $\omega$ (Fig. 6). The radius of solenoid is equal to $r$, the distance between the particle and axis of solenoid is $R>r$. The axis of solenoid is collinear to the axis $z$. Under calculation of the force, acting on the solenoid due to the charged particle, we assume that at $t=0$ the axis $x$ is orthogonal to orbital velocity of particle. Then, using designations of Fig. 5, we again obtain Eq. (A9). However, now $R>r$, and we derive for a single loop

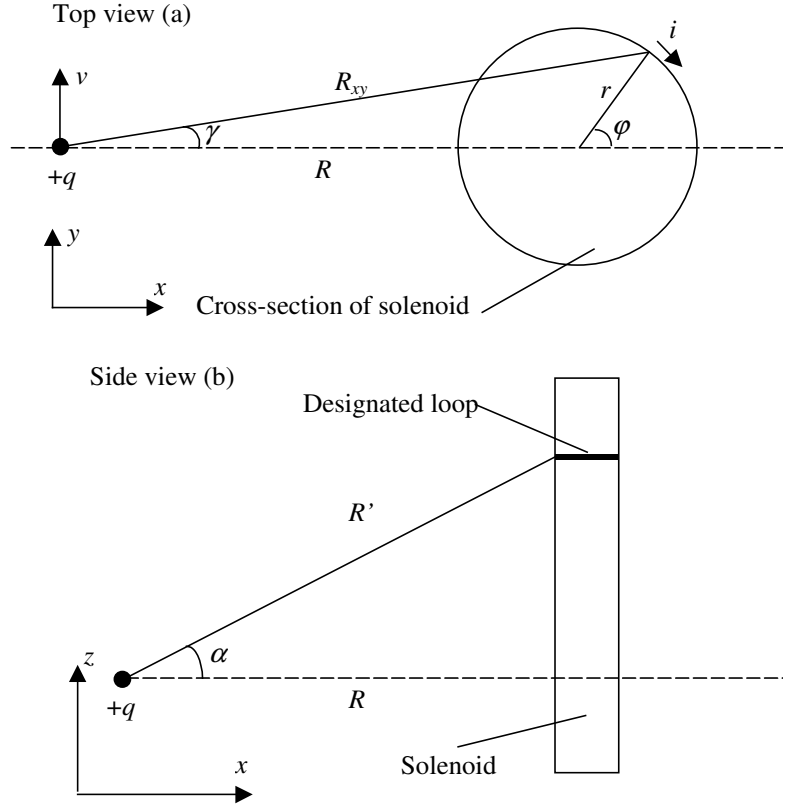

Fig. 6. The charged particle $+q$ orbits around the solenoid

$$dF_{1x} = \frac{qvir}{4\pi\varepsilon_0 c^2} \int_0^{2\pi} \frac{(R+r\cos\varphi)\cos\varphi\, d\varphi}{\left(R^2 + 2Rr\cos\varphi + r^2 + z^2\right)^{3/2}} = \frac{qvir}{4\pi\varepsilon_0 c^2 R} \int_0^{2\pi} \frac{\left(1+\frac{r}{R}\cos\varphi\right)\cos\varphi\, d\varphi}{\left(1 + \frac{2r}{R}\cos\varphi + \frac{r^2}{R^2} + \frac{z^2}{R^2}\right)^{3/2}}.$$

The fragment of solenoid with the length $dz$ contains $ndz$ loops. Hence, the force, acting on the fragment with the length $dz$ is

$$dF_x = \frac{qvirndz}{4\pi\varepsilon_0 c^2 R} \int_0^{2\pi} \frac{\left(1+\frac{r}{R}\cos\varphi\right)\cos\varphi\, d\varphi}{\left(1 + \frac{2r}{R}\cos\varphi + \frac{r^2}{R^2} + \frac{z^2}{R^2}\right)^{3/2}}.$$

From there the total force acting on the solenoid along the axis $x$ is

$$F_x = \frac{qvirn}{4\pi\varepsilon_0 c^2 R} \int_{-\infty}^{\infty}\int_0^{2\pi} \frac{\left(1+\frac{r}{R}\cos\varphi\right)\cos\varphi\, d\varphi dz}{\left(1 + \frac{2r}{R}\cos\varphi + \frac{r^2}{R^2} + \frac{z^2}{R^2}\right)^{3/2}} \quad \text{or}$$

$$F_x = \frac{qvrB}{4\pi R} \int_{-\infty}^{\infty}\int_0^{2\pi} \frac{\left(1+\frac{r}{R}\cos\varphi\right)\cos\varphi\, d\varphi dz}{\left(1 + \frac{2r}{R}\cos\varphi + \frac{r^2}{R^2} + \frac{z^2}{R^2}\right)^{3/2}}.$$

This equation can also be expressed via the value of vector potential of solenoid $A$, using the equality $A = Br^2/R$ (outside the solenoid):

$$F_x = \frac{qvA}{2\pi r} \int_{-\infty}^{\infty}\int_0^{2\pi} \frac{\left(1+\frac{r}{R}\cos\varphi\right)\cos\varphi\, d\varphi dz}{\left(1 + \frac{2r}{R}\cos\varphi + \frac{r^2}{R^2} + \frac{z^2}{R^2}\right)^{3/2}}. \tag{B1}$$

Integration over $z$ gives:

$$F_x = \frac{qvAR^2}{2\pi r}\int_0^{2\pi}\left(1+\frac{r}{R}\cos\varphi\right)\cos\varphi\, d\varphi \int_{-\infty}^{\infty}\frac{dz}{\left(R^2 + 2rR\cos\varphi + r^2 + z^2\right)^{3/2}} =$$

$$= \frac{qvAR^2}{\pi r}\int_{-\infty}^{\infty}\frac{\left(1+\frac{r}{R}\cos\varphi\right)}{R^2 + 2rR\cos\varphi + r^2}\cos\varphi\, d\varphi = \frac{qvA}{\pi r}\int_{-\infty}^{\infty}\frac{\left(1+\frac{r}{R}\cos\varphi\right)}{1+\frac{2r}{R}\cos\varphi + \frac{r^2}{R^2}}\cos\varphi\, d\varphi \tag{B2}$$

The remaining integral over $\varphi$ is equal to

$$\int_{-\infty}^{\infty}\frac{\left(1+\frac{r}{R}\cos\varphi\right)}{1+\frac{2r}{R}\cos\varphi + \frac{r^2}{R^2}}\cos\varphi\, d\varphi = -\frac{\pi r}{R}. \tag{B3}$$

Substituting Eq. (B3) into Eq. (B2), we obtain

$$F_x = -qvA/R. \tag{B4}$$

This expression describes the momentary force due to the rotating particle with the negative *x*-coordinate, when the axis *x* be orthogonal to its orbital velocity. Similar calculations show that the *y*- and *z*-components of force, defined by Eqs. (A4) and (A5), correspondingly, both equal to zero. Hence, Eq. (B4) describes the total momentary force due to the rotating particle with the negative *x*-coordinate, when the axis *x* be orthogonal to its orbital velocity. It shows that the force is directed along the line, joining the axis of solenoid and momentary position of the rotating particle. One follows from there that the direction of the force, exerted by the particle on solenoid, rotates together with the particle at the same angular frequency $\omega$. Hence, the projections of this force change with time for a laboratory observer as

$$F_x = qvA/R = q\omega A \cos \omega t. \tag{B5}$$

$$F_y = qvA/R = q\omega A \sin \omega t, \tag{B6}$$

and $F_z = 0$.

One can see that Eqs. (B5) and (B6), taken together, can be written in the vector form as

$$\vec{F} = -q(\vec{\omega} \times \vec{A}). \tag{B7}$$

For the vector field of solenoid, circulated in the clock-wise direction, we write

$$\frac{d\vec{A}}{dt} = (\vec{\omega} \times \vec{A}). \tag{B8}$$

Comparison of Eqs. (B7) and (B8) shows that

$$\vec{F} = -q\,d\vec{A}/dt = -d\vec{P}_A/dt. \tag{B9}$$

Thus, we have shown that the force, acting on the solenoid due to a rotating particle, is equal with the opposite sign to the total time derivative of the potential momentum $q\vec{A}$ for the system "solenoid +particle".

**Appendix C. The energy flux in electromagnetic field of a straight wire with a steady current**

It is known that the energy flux of a steady current, when determined through the Poynting vector, is orthogonal to the wire (Fig. 7).

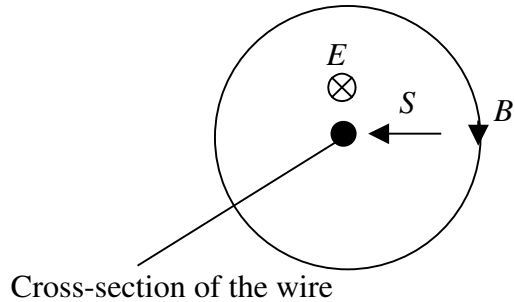

Fig. 7. The current enter into the sheet. The directions of electric field $\vec{E}$, magnetic field $\vec{B}$ and Poynting vector $\vec{S}$ are indicated.

In general, this result seems puzzling, because it would be naturally to expect, that conduction electrons take their energy from the source of electric field, but not from surrounding space. However, till the moment the result was considered as a reflection of strange properties of energy fluxes in bound EM fields, which contradict our intuition [1]. Now we will show that our intuition is right, while the strange energy flux resulted from an incorrect writing of the energy balance equation for the problem considered. Namely, the time rate of heat energy $\dot{H}$, dissipating in the wire by flowing current, should be explicitly included. Since

$$\dot{H} = \vec{j} \cdot \vec{E}, \qquad (C1)$$

then Eq. (5) transforms into

$$\frac{\partial u}{\partial t} + \nabla \cdot \vec{S} + \dot{H} = 0. \qquad (C2)$$

However, the energy density of EM field cannot be changed due to the heat energy. Hence, in order to determine a true direction of the energy flux of EM field, now we have to exclude $\dot{H}$ from Eq. (C2). It can be done with the equation

$$\nabla \cdot \vec{S} + \vec{E} \cdot \vec{j} = \nabla \vec{S}_U, \qquad (C3)$$

following under comparison of Eqs. (5) and (52). Then, combining Eqs. (C1)-(C3), we obtain

$$\frac{\partial u}{\partial t} + \nabla \cdot \vec{S}_U = 0,$$

where $\nabla \cdot \vec{S}_U = u\vec{v}$, $\vec{v}$ being the velocity of carriers of current. Thus, we find that the energy flux is parallel to the direction of current and its velocity coincides with the flow velocity.

Finally, we mention that Eq. (C3) gives a key to understand, why Eq. (8) can be successfully applicable for description of a momentum of bound EM field. However, this problem falls outside the scope of the present paper.